\documentclass[review,3p,authoryear,times]{elsarticle}
\usepackage{color}
\usepackage{multirow,setspace,times,amssymb,amsmath,graphicx,color,subfigure,url}
\usepackage{dsfont,bm}
\usepackage{threeparttable}
\usepackage{rotating}
\usepackage{lscape,pdflscape}
\usepackage{dcolumn}
\usepackage{listings}
\usepackage{relsize}
\usepackage{booktabs}
\usepackage[colorlinks=true,linkcolor=blue,citecolor=blue,urlcolor=blue]{hyperref}
\usepackage{tikz}
\usepackage{caption}
\usepackage{pdflscape}
\usepackage{float}
\usepackage{afterpage}     
\usepackage{placeins}      
\usepackage{lipsum}        
\usepackage[capitalize]{cleveref} 
\usepackage{float}
\usepackage{pdfpages}
\journal{Elsevier}

\begin{document}
	
	\begin{frontmatter}
		
		\title{Contemporaneous and lagged spillovers  between agriculture, crude oil, carbon emission allowance, and climate change}
		
		\author[SILC]{Yan-Hong Yang }

		\author[SUIBE]{Ying-Hui Shao}
		\ead{yinghuishao@126.com}
		
		\author[SB,RCE,SM]{Wei-Xing Zhou \corref{cor1}}
		\ead{wxzhou@ecust.edu.cn}
		\cortext[cor1]{Corresponding author. }
		
		\address[SILC]{SILC Business School, Shanghai University, Shanghai 201899, China}
		\address[SUIBE]{School of Statistics and Information, Shanghai University of International Business and Economics, Shanghai 201620, China}
		\address[SB]{School of Business, East China University of Science and Technology, Shanghai 200237, China}
		\address[RCE]{Research Center for Econophysics, East China University of Science and Technology, Shanghai 200237, China}
		\address[SM]{School of Mathematics, East China University of Science and Technology, Shanghai 200237, China}
		
		\begin{abstract}
			In this paper, we examine the contemporaneous and lagged spillovers among the agricultural, crude oil, carbon emission allowance, and climate change markets. Adopting the $R^2$ decomposed connectedness approach, our empirical analysis reveals several key findings. First, the overall total connectedness index (TCI) dynamics have been mainly dominated by contemporaneous effects. Second, there are heterogeneous spillover effects among agricultural markets. Specially, corn, soybean meal, and wheat are the major risk transmitters to this system, while barley, cocoa, and lean hog are the main risk receivers of shocks. Third, we also find that climate change has significant spillovers to other markets.
		\end{abstract}
		
	\end{frontmatter}

	\section{Introduction}
	\label{S1:Introduction}
	
	The agricultural market is crucial to the national economy, which is susceptible to external shocks. Intensifying global warming and extreme weather have heightened attention to climate risks \citep{venturini2022climate,kemp2022climate,pham2024blessings,khalfaoui2024impact}, severely impacting crop cultivation. Additionally, carbon dioxide also exerts a negative impact on agricultural markets \citep{khalfaoui2024impact}. Agricultural and energy markets have long been connected  \citep{du2011speculation}, with climate change adding complexity by affecting fossil energy extraction and transport. Volumetric oil production declines with significant growth of greenhouse gas \citep{masnadi2017climate}.  However, energy production and agriculture can worsen pollution, climate conditions, and global warming \citep{jin2023geopolitical}.

	Owing to the strong interconnections within the financial system, a shock in one sector can affect other sectors \citep{wang2020identifying,wang2022measuring}. Furthermore, the contagion effect can spread across the financial market, making spillover research among agricultural, energy, and other markets vital for assessing risks, price forecasting, and portfolio management \citep{wei2023alarming}. \citet{jin2023geopolitical} report the dynamic spillover among geopolitical risk, climate risk, and energy markets from an international perspective. \citet{lin2024stress} observe simultaneous opposite correlations between climate change attention and crude oil markets.

	In response to the crucial climate issues, the  carbon emission allowance trading market has been developed, helping to mitigate climate change. It should be noted that fossil energy and agricultural markets have important impacts on the carbon market, which will influence carbon demand, supply, and prices \citep{gong2021analyzing}. \citet{chen2024dynamic} find that the risk is transferred from the international gas and crude oil markets to the Chinese carbon market. \citet{younis2024exploring} invesitgate contagion risk among climate change, renewable energy, technological innovation, and G-17 banking stock markets. They find significant risk spillover and connectedness.

	There are various measures to investigate risk transmission in agricultural and other commodity markets, including DCC-GARCH \citep{luo2018high}, CoVaR \citep{kumar2021time}, TVP-VAR-DY \citep{zhu2024uncovering}, quantile VAR model \citep{tiwari2022quantile,wei2023alarming}. Recently, \cite{balli2023contemporaneous} and \citet{naeem2024measuring} introduce the $R^2$ decomposed connectedness approach, efficiently splitting connectedness into contemporaneous and lagged components while ensuring data standardization. This method has since been applied to risk spillover analysis in financial markets \citep{li2024assessing,liu2024decomposing}.

	Commodity financialization boosts liquidity and market efficiency, but also raises contagion risk  \citep{goldstein2022commodity,bianchi2020financialization}. Hence, we examine the spillover among agriculture, crude oil, carbon emission allowance, and climate change futures markets via the $R^2$ decomposed connectedness approach. Our study contributes the following: First, we apply the $R^{2}$ decomposed connectedness to accurately identify the immediate and delayed information transmissions between commodity markets, providing a comprehensive understanding of how risks are transmitted across different time dimensions. Second, this paper deepens the understanding of risk transmission between commodity markets by expanding the study on factors influencing commodity prices. Third, we conduct detailed analysis on how different agricultural products respond to other futures fluctuations, offering targeted risk management strategies for investors and policymakers.

	The paper is organized as follows: Section~\ref{S2:Methodology} outlines the methodology, Section~\ref{S3:Data description} describes the data, Section~\ref{S4:Empirical analysis} presents empirical results and robustness checks, and Section~\ref{S5:conclusion} concludes.

	\section{Methodology}
	\label{S2:Methodology}
	
	Building on the $R^2$ decomposed connectedness approach \citep{balli2023contemporaneous,naeem2024measuring}, this study identifies both contemporaneous and lagged idiosyncratic spillovers in agricultural and other futures markets, expanding earlier research that focused solely on contemporaneous effects. As detailed by \cite{balli2023contemporaneous}, the $R^2$ decomposed connectedness framework utilizes a VAR($p$) model to incorporate contemporaneous effects:
	
	\begin{equation}
	\bm{y}_t = \sum_{i=0}^{p} \bm{B}_i \bm{y}_{t-i} + \bm{u}_t, \quad \bm{u}_t \sim N(\bm{0}, \bm{\Sigma})
	\label{Eq:VAR}
	\end{equation}
	where \(\bm{y}_t\) and \(\bm{u}_t\) denote demeaned \(K \times 1\) dimensional vectors over time, with \(\bm{B}\) and \(\bm{\Sigma}\) being dimensional matrices  such that $diag(\bm{B}_0) = 0$. Notably, when \(p\) is set to zero, the model aligns with the contemporaneous \(R^2\) decomposed connectedness framework as introduced by \cite{naeem2024measuring}. In other words, Eq.~(\ref{Eq:VAR}) can be delineated as \(\bm{y}_{k,t} = \bm{b}_k\bm{x}_k + \bm{u}_{k,t}\), where \(\bm{x}_t = [\bm{y}_{t},\bm{y}_{t-1}, \ldots, \bm{y}_{t-i}, \ldots, \bm{y}_{t-p}]\) constitutes an \(K(p + 1) \times 1\) dimensional vector, and \(\bm{b}_k\) is structured as an \(1 \times K(p + 1)\) dimensional vector with its $k$th component being zero.

	Typically, the cumulative \( R^2 \) contributions from bivariate linear regressions (BLRs) correspond with the multivariate linear regression (MLR)'s goodness-of-fit only when the variables on the right-hand side (RHS) are not correlated with one another. It is crucial to establish a transformation that converts the correlated series \( \bm{x}_{k,t} \)\footnote{The vector \( \bm{x}_{k,t} \) corresponds to \( \bm{x}_{t} \)but omits the variable on the left-hand side.} into a set of orthogonal series. This transformation is facilitated by implementing principal component analysis, where the number of latent factors matches the quantity of RHS variables. As a result, the \( R^2 \) decomposition of an MLR can be formulated as follows:
	
	\begin{equation}
	\bm{R}_{xx}=\bm{V}\bm{\Lambda} \bm{V}^{\prime}=\bm{C}\bm{C}^{\prime}
	\end{equation}
	
	\begin{equation}
	\bm{C} = \bm{V} \bm{\Lambda}^{1/2} \bm{V}^{\prime}
	\end{equation}
	
	\begin{equation}
	\bm{R}^{2,d} = \bm{C}^2(\bm{C}^{-1}\bm{R}_{yx})^2
	\end{equation}
	where \( \bm{V} \), \( \bm{\Lambda} = diag(\lambda_1, ..., \lambda_{K(p+1)-1}) \), alongside \( \bm{R}_{xx} \), constitute the eigenvectors, eigenvalues, and Pearson correlation matrices of dimensions \( [K(p+1)-1] \times [K(p+1)-1] \) respectively. Additionally, \( \bm{R}_{yx} \) and the \( \bm{R}^{2,d} \) vectors provide Pearson correlation and \( R^2 \) contribution metrics, each of dimension \( [K(p + 1) - 1] \times 1 \). To elaborate, \( \bm{R}_{xx} \) captures the Pearson correlation across the RHS variables, whereas \( \bm{R}_{yx} \) details the correlation between LHS and RHS variables. The initial set of \( K - 1 \) values in \( \bm{R}^{2,d} \) corresponds to the contemporaneous \( R^2 \) contributions, and the subsequent values identify the  lagged \( R^2 \) contributions. Thus, the sum of the \( \bm{R}^{2,d} \) vector is tantamount to the MLR’s overall goodness-of-fit measure. Finally, stacking the \( \bm{R}^{2,d} \) vectors from all $K$ multivariate regressions yields a \( K \times K(p + 1) \) dimensional \( \bm{R}^{2,d} \) decomposition matrix, detailed as \( [\bm{R}^{2,d}_0; ...; \bm{R}^{2,d}_i; ... \bm{R}^{2,d}_p] \).
	Consequently, the component \( \bm{R}^{2,d}_0 \)\footnote{The diagonal entries of the matrix \( \bm{R}^{2,d}_0 \) are all zero.} signifies the contemporaneous spillover effects \( \bm{R}^{2,d}_C \), whereas the sum \( \bm{R}^{2,d}_1 + \ldots + \bm{R}^{2,d}_i + \ldots + \bm{R}^{2,d}_p \) represents the aggregate of lagged spillover effects \( \bm{R}^{2,d}_L \). 
	
	Within the \cite{diebold2014network} framework for measuring connectedness, these components serve as substitutes for the scaled GFEVD matrix, indicating that the total connectedness index (TCI) is the mean \( R^2 \) across all multivariate linear regressions.

	\begin{equation}
	TCI=\frac{1}{K}\sum_{k=1}^{K}R_{k}^{2}=\left(\frac{1}{K}\sum_{k=1}^{K}\sum_{i=1}^{K}\bm{R}_{C,k,i}^{2,d}\right)+\left(\frac{1}{K}\sum_{k=1}^{K}\sum_{i=1}^{K}\bm{R}_{L,k,i}^{2,d}\right)=TCI^{C}+TCI^{L}
	\end{equation}
	here, \(TCI^C\) and \(TCI^L\) denote the contemporaneous and lagged total connectedness indices, respectively. In addition, since the goodness-of-fit $R_{k}^{2}$ measure ranges from 0 to 1, there is no need for normalization techniques \citep{chatziantoniou2021interest}.
	Further, total directional connectedness TO and FROM others, alongside the net total directional connectedness, are expressed as:

	\begin{equation}
	TO_{i}=\sum_{k=1}^{K}\bm{R}_{C,k,i}^{2,d}+\sum_{k=1}^{K}\bm{R}_{L,k,i}^{2,d}=TO_i^C+TO_i^L
	\end{equation}

	\begin{equation}
	FROM_{i}=\sum_{k=1}^{K}\bm{R}_{C,i,k}^{2,d}+\sum_{k=1}^{K}\bm{R}_{L,i,k}^{2,d}=FROM_i^C+FROM_i^L
	\end{equation}

	\begin{equation}
	\begin{split}
	NET_i &= TO_i - FROM_i \\
	&= (TO_i^C+TO_i^L)-(FROM_i^C+FROM_i^L) \\
	&= (TO_i^C-FROM_i^C)+(TO_i^L-FROM_i^L) \\
	&= NET^{C}_i + NET^{L}_i.
	\end{split}
	\end{equation}
	The $TO_i$ index quantifies the variance in LHS variables due to $i$, while the $FROM_i$ index measures how much RHS variables explain the variance in $i$. The $NET_i$ index shows whether $i$ is a net transmitter or receiver of shocks.	If series \( i \) has a positive (negative) $NET$ value, it is deemed a net transmitter (receiver) of shocks, suggesting it influences other variables more (less) than it is influenced by them. Similarly, net pairwise directional connectedness ($NPDC_{ij}$) operates like $NET$ but between pairs. A positive (negative) $NPDC_{ij}$ means $j$ explains more (less) of $i$'s variation than vice versa.

	\section{Data}
	\label{S3:Data description}

\begin{figure}[t]
		\centering
		\includegraphics[width=1\linewidth]{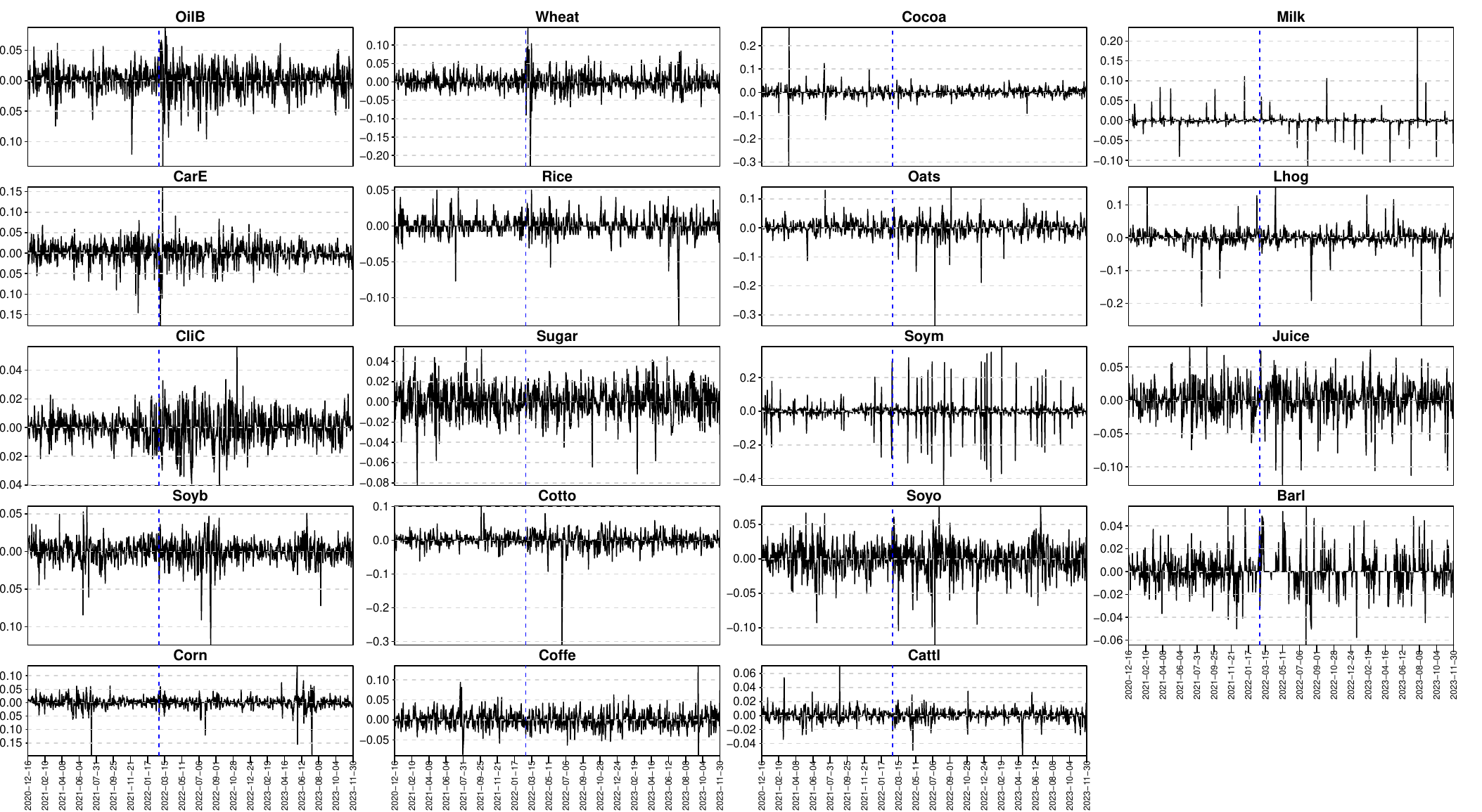}
		\caption{Returns.}
		\label{Fig:Returns}
	\end{figure}

	Our dataset includes 19 futures contracts in agriculture, crude oil, carbon emission allowances, and climate change, spanning from December 15, 2020, to November 30, 2023, with 773 daily closing price observations. The starting date, December 15, 2020, marks the launch of climate change futures. Missing data points are filled with the previous day's data for consistency. The data is sourced from Wind  
	and Bloomberg. 
	We compute logarithmic returns of futures prices as \( \ln\left(\frac{p_{i,t}}{p_{i,t-1}}\right) \), where \( p_{i,t} \) is the daily closing price of the \( i \)-th future at time \( t \), as shown in Fig.~\ref{Fig:Returns}.
	
		We include Brent Oil Futures from the ICE Exchange in the oil category. For carbon emission allowance, our dataset features the EUA Carbon traded on ICE Futures Europe. In the climate segment, we include the MSCI World Climate Change Futures from the ICE Exchange, a key addition given the global focus on climate-related financial products.
		
			Our dataset includes a wide range of agricultural futures: Soybean, Soybean meal, Soybean oil, Corn, Wheat, Oats, and Rough Rice, all traded on the CBOT Exchange. The ICE Exchange is represented with Sugar \#11 Futures, Cotton \#2 Futures, Orange juice, Cocoa, and Coffee C Futures; notably, the Coffee C futures contract on ICE serves as the global benchmark for Arabica coffee. Additionally, Live cattle, Milk, 
	and Lean hog futures from the CME 
	Exchange  are part of our selection. Given Australia's significance in the global barley market, we include Barley Futures from the Australian Securities Exchange.
	
	\begin{table}[!htbp]
		\centering
		\begin{threeparttable}
			\small
			\caption{Summary statistics.}
			\label{Tb:Statistics}
			\begin{tabular}{lllllllllllllllllll}
				\toprule
				& Mean & Variance & Skewness & Ex.Kurtosis & JB & ERS & ADF & PP \\
				\midrule
				OilB & 0.001 & 0.001 & -0.730*** & 3.370*** & 433.988*** & -10.328*** & -10.645*** & -678.469*** \\
				CarE & 0.001 & 0.001 & -0.577*** & 5.491*** & 1012.642*** & -8.316*** & -9.636*** & -817.237*** \\
				CliC & 0.000 & 0.000 & -0.020 & 1.779*** & 101.843*** & -9.559*** & -8.786*** & -683.829*** \\
				Soyb & 0.000 & 0.000 & -1.095*** & 7.565*** & 1994.993*** & -12.933*** & -8.913*** & -723.122*** \\
				Corn & 0.000 & 0.001 & -1.898*** & 19.347*** & 12503.623*** & -12.054*** & -8.895*** & -810.834*** \\
				Wheat & 0.000 & 0.001 & -0.358*** & 11.338*** & 4151.900*** & -11.440*** & -8.625*** & -768.58*** \\
				Rice & 0.000 & 0.000 & -1.354*** & 15.520*** & 7983.829*** & -12.706*** & -10.135*** & -625.387*** \\
				Sugar & 0.001 & 0.000 & -0.281*** & 1.780*** & 112.122*** & -5.708*** & -10.479*** & -760.936*** \\
				Cotto & 0.000 & 0.001 & -3.375*** & 46.721*** & 71680.628*** & -13.173*** & -8.824*** & -720.521*** \\
				Coffe & 0.001 & 0.000 & 0.433*** & 2.946*** & 303.292*** & -7.954*** & -8.99*** & -761.17*** \\
				Cocoa & 0.001 & 0.001 & -1.083*** & 71.817*** & 166056.293*** & -1.935* & -11.786*** & -801.937*** \\
				Oats & 0.000 & 0.001 & -2.631*** & 31.460*** & 32726.651*** & -13.461*** & -9.374*** & -665.873*** \\
				Soym & 0.000 & 0.005 & -0.459*** & 13.158*** & 5596.583*** & -13.324*** & -12.271*** & -917.347*** \\
				Soyo & 0.000 & 0.001 & -0.653*** & 2.701*** & 289.566*** & -11.401*** & -7.842*** & -672.379*** \\
				Cattl & 0.001 & 0.000 & 0.283*** & 11.247*** & 4079.262*** & -11.856*** & -8.694*** & -767.425*** \\
				Milk & 0.000 & 0.000 & 3.442*** & 68.055*** & 150503.046*** & -8.076*** & -8.876*** & -779.297*** \\
				Lhog & 0.000 & 0.001 & -2.205*** & 32.473*** & 34545.979*** & -11.575*** & -9.838*** & -754.072*** \\
				Juice & 0.002 & 0.001 & -0.762*** & 3.566*** & 483.635*** & -5.442*** & -11.262*** & -678.677*** \\
				Barl & 0.000 & 0.000 & 0.273*** & 4.315*** & 608.492*** & -9.041*** & -7.717*** & -884.229*** \\
				\bottomrule
			\end{tabular}
			\begin{tablenotes}
				\footnotesize
				\item Notes: JB represents the Jarque-Bera normality test statistic; ERS, ADF, and PP show the results of the unit root tests. \textit{*} indicates 10\% significance level, \textit{**} indicates 5\% significance level, \textit{***} indicates 1\% significance level.
			\end{tablenotes}
		\end{threeparttable}
	\end{table}

	To simplify the presentation of tables and figures, we employ the abbreviations OilB, CarE, CliC, Soyb, Rice, Sugar, Cotto, Coffe, Soym, Soyo, Cattl, Lhog, Juice, and Barl to denote the futures Brent Oil, EUA Carbon, MSCI World Climate Change, Soybean, Rough Rice, Sugar \#11, Cotton \#2, Coffee, Soybean Meal, Soybean Oil, Live Cattle, Lean Hog, Orange Juice, and Barley, respectively.
	
	\begin{figure}[htp]
		\centering
		\includegraphics[width=1\linewidth]{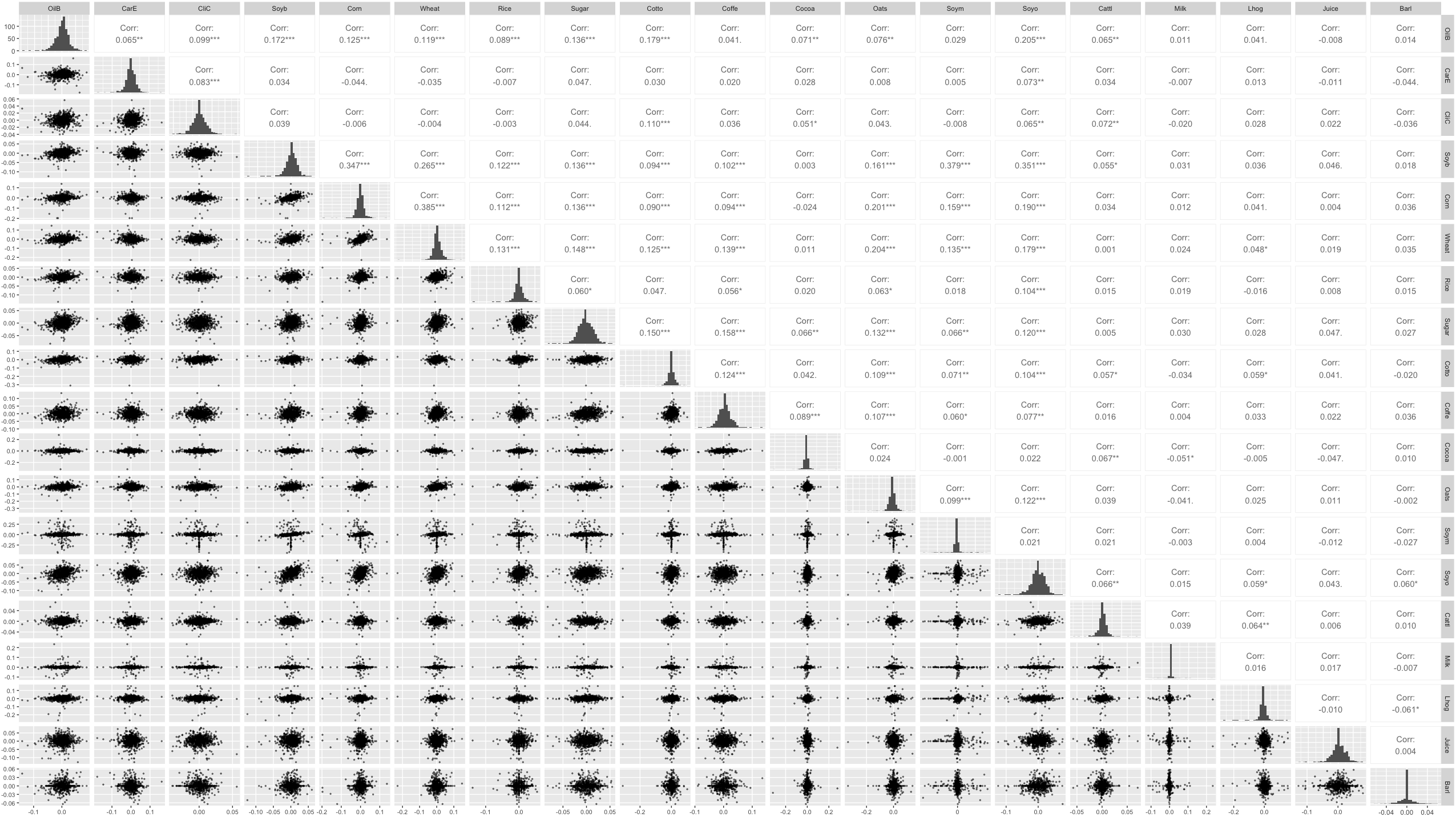}
		\caption{Kendall correlations. Notes: * Significant at 10\%; ** significant at 5\%; *** significant at 1\%.
		}
		\label{Fig:Correlation}
	\end{figure}
	
	Table~\ref{Tb:Statistics} reports the descriptive statistics. Except coffee, cattle, milk, and barley, other variables are left skewed. Besides that, with the exception of climate change, sugar, and soybean oil, others are leptokurtic.The JB, ERS, ADF, and PP tests indicate that all series are non-normally distributed and stationary. Moreover, the Kendall rank correlation coefficients presented in Fig.~\ref{Fig:Correlation}   suggest that most variables are positively correlated.

	\section{Empirical analysis}
	\label{S4:Empirical analysis}
	
	\subsection{Averaged dynamic connectedness measures}
	
	Table~\ref{Tb:Average:Dynamic:Connectedness} shows the averaged connectedness measures, with parentheses indicating contemporaneous and lagged $R^{2}$ decomposed values. The average TCI is 29.38\%, with 20.43\% from contemporaneous effects and 8.95\% from lagged dynamics, indicating that 29.38\% of the LHS variables' variance is explained by the RHS variables. 
	
	Furthermore, almost all the contemporaneous FROM and TO connectedness metrics are significantly greater than the corresponding lagged counterparts. However, cocoa, milk, and barley have lagged FROM measures (11.03\%, 9.81\%, and 13.34\%) exceeding their 
	FROM contemporaneous measures (10.28\%, 7.65\%, and 10.84\%). The lagged TO measures for cattle and milk (10.14\% and 7.29\%, respectively) are both higher than their contemporaneous values (9.45\% and 6.89\%, respectively). 
	
	\begin{table}[H]
		\caption{Averaged dynamic connectedness.}
		\label{Tb:Average:Dynamic:Connectedness}
		\resizebox{\textwidth}{!}{
			\centering	
			\begin{tabular}{lllllllllllllllllllll}
				\toprule
				& OilB & CarE & CliC & Soyb & Corn & Wheat & Rice & Sugar & Cotto & Coffe & Cocoa & Oats & Soym & Soyo & Cattl & Milk & Lhog & Juice & Barl & FROM\\
				\midrule
				OilB & 0.20 & 2.58 & 3.13 & 3.86 & 2.14 & 1.89 & 1.34 & 3.24 & 3.44 & 1.07 & 1.03 & 1.25 & 1.75 & 4.78 & 0.81 & 1.27 & 1.02 & 1.20 & 0.97 & 36.76\\
				& ( 0.00,  0.20) & ( 1.34,  1.23) & ( 2.59,  0.54) & ( 3.52,  0.34) & ( 1.68,  0.47) & ( 1.17,  0.72) & ( 1.16,  0.18) & ( 2.58,  0.66) & ( 3.29,  0.15) & ( 0.75,  0.32) & ( 0.68,  0.35) & ( 0.60,  0.65) & ( 0.68,  1.07) & ( 4.35,  0.42) & ( 0.40,  0.42) & ( 0.81,  0.46) & ( 0.29,  0.73) & ( 0.18,  1.02) & ( 0.63,  0.34) & (26.70, 10.06)\\
				CarE & 1.91 & 1.23 & 2.12 & 0.83 & 1.62 & 1.87 & 1.28 & 1.75 & 0.80 & 0.93 & 1.23 & 1.02 & 0.45 & 1.16 & 0.86 & 0.57 & 0.57 & 1.12 & 1.07 & 21.15\\
				& ( 1.45,  0.46) & ( 0.00,  1.23) & ( 1.49,  0.63) & ( 0.56,  0.28) & ( 1.23,  0.39) & ( 1.57,  0.30) & ( 0.49,  0.78) & ( 1.44,  0.31) & ( 0.47,  0.33) & ( 0.20,  0.73) & ( 0.46,  0.77) & ( 0.38,  0.63) & ( 0.17,  0.28) & ( 0.93,  0.23) & ( 0.66,  0.19) & ( 0.26,  0.31) & ( 0.43,  0.14) & ( 0.72,  0.40) & ( 0.89,  0.18) & (13.80,  7.35)\\
				CliC & 3.10 & 1.84 & 0.47 & 0.74 & 0.80 & 0.53 & 0.57 & 1.62 & 2.35 & 1.05 & 0.58 & 0.78 & 0.64 & 1.77 & 1.55 & 1.15 & 0.97 & 1.11 & 0.92 & 22.07\\
				& ( 2.77,  0.33) & ( 1.51,  0.33) & ( 0.00,  0.47) & ( 0.43,  0.31) & ( 0.33,  0.47) & ( 0.30,  0.23) & ( 0.19,  0.38) & ( 0.59,  1.02) & ( 2.12,  0.24) & ( 0.57,  0.49) & ( 0.38,  0.20) & ( 0.55,  0.23) & ( 0.47,  0.17) & ( 0.68,  1.08) & ( 0.89,  0.67) & ( 0.59,  0.56) & ( 0.43,  0.54) & ( 0.54,  0.57) & ( 0.42,  0.50) & (13.76,  8.31)\\
				Soyb & 3.51 & 0.52 & 0.60 & 0.38 & 8.91 & 4.13 & 1.89 & 1.51 & 1.19 & 1.08 & 0.73 & 3.80 & 5.29 & 16.29 & 1.21 & 0.68 & 3.38 & 1.26 & 0.73 & 56.71\\
				& ( 3.33,  0.18) & ( 0.42,  0.10) & ( 0.33,  0.27) & ( 0.00,  0.38) & ( 8.35,  0.56) & ( 3.60,  0.53) & ( 1.54,  0.35) & ( 0.93,  0.58) & ( 0.51,  0.68) & ( 0.74,  0.34) & ( 0.56,  0.17) & ( 1.58,  2.22) & ( 3.65,  1.64) & (15.04,  1.25) & ( 0.46,  0.75) & ( 0.11,  0.57) & ( 2.70,  0.67) & ( 1.05,  0.22) & ( 0.48,  0.25) & (45.39, 11.32)\\
				Corn & 2.26 & 1.19 & 0.49 & 9.46 & 1.98 & 9.42 & 2.70 & 1.06 & 1.03 & 0.60 & 0.61 & 2.67 & 6.71 & 3.31 & 0.75 & 0.37 & 0.69 & 1.28 & 1.54 & 46.14\\
				& ( 1.66,  0.60) & ( 1.07,  0.12) & ( 0.25,  0.23) & ( 8.91,  0.55) & ( 0.00,  1.98) & ( 8.71,  0.70) & ( 2.42,  0.28) & ( 0.56,  0.50) & ( 0.85,  0.18) & ( 0.41,  0.20) & ( 0.42,  0.19) & ( 2.31,  0.36) & ( 5.74,  0.97) & ( 2.88,  0.43) & ( 0.24,  0.51) & ( 0.26,  0.12) & ( 0.43,  0.26) & ( 0.88,  0.41) & ( 1.38,  0.17) & (39.37,  6.77)\\
				Wheat & 1.97 & 1.71 & 0.61 & 4.30 & 10.01 & 0.64 & 3.74 & 3.55 & 1.11 & 1.44 & 1.66 & 3.96 & 0.92 & 2.50 & 1.07 & 0.80 & 1.21 & 0.44 & 1.13 & 42.12\\
				& ( 1.16,  0.81) & ( 1.39,  0.32) & ( 0.25,  0.36) & ( 3.79,  0.51) & ( 8.97,  1.04) & ( 0.00,  0.64) & ( 3.09,  0.65) & ( 3.19,  0.36) & ( 0.64,  0.47) & ( 0.90,  0.54) & ( 0.32,  1.35) & ( 3.72,  0.23) & ( 0.68,  0.24) & ( 1.97,  0.53) & ( 0.47,  0.60) & ( 0.29,  0.51) & ( 0.74,  0.46) & ( 0.29,  0.16) & ( 0.74,  0.38) & (32.60,  9.52)\\
				Rice & 2.05 & 0.59 & 0.44 & 2.56 & 3.71 & 3.61 & 2.61 & 1.75 & 0.93 & 0.88 & 1.06 & 1.98 & 2.08 & 1.78 & 1.43 & 0.67 & 0.47 & 0.55 & 0.57 & 27.11\\
				& ( 1.21,  0.85) & ( 0.47,  0.12) & ( 0.18,  0.26) & ( 1.85,  0.71) & ( 2.68,  1.04) & ( 3.32,  0.29) & ( 0.00,  2.61) & ( 1.56,  0.18) & ( 0.75,  0.19) & ( 0.56,  0.32) & ( 0.78,  0.28) & ( 1.54,  0.44) & ( 0.65,  1.43) & ( 0.86,  0.91) & ( 0.14,  1.29) & ( 0.28,  0.39) & ( 0.20,  0.28) & ( 0.22,  0.33) & ( 0.20,  0.37) & (17.43,  9.68)\\
				Sugar & 2.86 & 1.67 & 1.19 & 1.92 & 0.77 & 3.84 & 1.91 & 0.35 & 1.75 & 3.97 & 1.48 & 3.16 & 0.54 & 1.33 & 1.70 & 1.10 & 0.55 & 1.05 & 0.54 & 31.34\\
				& ( 2.60,  0.26) & ( 1.36,  0.32) & ( 0.56,  0.64) & ( 0.94,  0.99) & ( 0.56,  0.22) & ( 3.36,  0.48) & ( 1.54,  0.37) & ( 0.00,  0.35) & ( 1.35,  0.40) & ( 3.68,  0.29) & ( 0.95,  0.53) & ( 2.77,  0.39) & ( 0.27,  0.27) & ( 1.02,  0.31) & ( 1.16,  0.54) & ( 0.31,  0.79) & ( 0.18,  0.37) & ( 0.52,  0.53) & ( 0.43,  0.11) & (23.53,  7.81)\\
				Cotto & 3.74 & 0.60 & 2.36 & 0.88 & 1.16 & 0.94 & 1.28 & 1.77 & 0.67 & 2.06 & 0.89 & 2.03 & 0.47 & 1.00 & 1.06 & 1.39 & 0.77 & 1.08 & 1.01 & 24.50\\
				& ( 3.45,  0.29) & ( 0.46,  0.14) & ( 2.10,  0.25) & ( 0.53,  0.35) & ( 0.94,  0.22) & ( 0.67,  0.27) & ( 0.76,  0.53) & ( 1.39,  0.38) & ( 0.00,  0.67) & ( 1.48,  0.58) & ( 0.39,  0.50) & ( 1.85,  0.18) & ( 0.23,  0.24) & ( 0.74,  0.26) & ( 0.36,  0.70) & ( 1.08,  0.31) & ( 0.62,  0.16) & ( 0.79,  0.29) & ( 0.28,  0.73) & (18.12,  6.38)\\
				Coffe & 2.23 & 0.43 & 0.93 & 1.26 & 0.82 & 1.31 & 0.79 & 4.34 & 1.83 & 0.33 & 1.75 & 1.95 & 0.44 & 0.67 & 0.72 & 0.73 & 1.22 & 0.94 & 0.54 & 22.92\\
				& ( 0.80,  1.43) & ( 0.19,  0.24) & ( 0.54,  0.39) & ( 0.89,  0.37) & ( 0.47,  0.35) & ( 0.97,  0.34) & ( 0.57,  0.22) & ( 3.80,  0.54) & ( 1.45,  0.38) & ( 0.00,  0.33) & ( 1.38,  0.37) & ( 1.38,  0.57) & ( 0.15,  0.29) & ( 0.33,  0.34) & ( 0.57,  0.15) & ( 0.50,  0.24) & ( 0.56,  0.66) & ( 0.71,  0.23) & ( 0.41,  0.13) & (15.68,  7.24)\\
				Cocoa & 1.19 & 1.36 & 1.39 & 1.15 & 1.57 & 1.12 & 1.57 & 1.42 & 0.80 & 1.83 & 1.09 & 0.81 & 0.83 & 1.78 & 1.18 & 0.28 & 1.46 & 0.84 & 0.71 & 21.30\\
				& ( 0.74,  0.44) & ( 0.46,  0.90) & ( 0.38,  1.01) & ( 0.71,  0.44) & ( 0.56,  1.01) & ( 0.38,  0.75) & ( 0.82,  0.75) & ( 1.01,  0.42) & ( 0.41,  0.39) & ( 1.39,  0.44) & ( 0.00,  1.09) & ( 0.22,  0.59) & ( 0.24,  0.60) & ( 0.94,  0.84) & ( 0.40,  0.78) & ( 0.11,  0.17) & ( 0.48,  0.98) & ( 0.69,  0.14) & ( 0.34,  0.37) & (10.28, 11.03)\\
				Oats & 1.06 & 0.68 & 1.59 & 2.06 & 3.50 & 4.37 & 1.98 & 3.01 & 2.37 & 1.75 & 0.31 & 0.67 & 0.92 & 4.72 & 0.47 & 0.86 & 0.48 & 1.60 & 1.01 & 32.75\\
				& ( 0.64,  0.42) & ( 0.34,  0.34) & ( 0.53,  1.07) & ( 1.87,  0.19) & ( 2.51,  0.99) & ( 3.93,  0.44) & ( 1.49,  0.49) & ( 2.74,  0.27) & ( 1.75,  0.61) & ( 1.29,  0.46) & ( 0.21,  0.10) & ( 0.00,  0.67) & ( 0.51,  0.41) & ( 4.36,  0.36) & ( 0.35,  0.13) & ( 0.36,  0.50) & ( 0.28,  0.20) & ( 0.57,  1.03) & ( 0.74,  0.27) & (24.46,  8.28)\\
				Soym & 0.92 & 0.36 & 0.60 & 5.05 & 6.92 & 1.10 & 0.82 & 0.48 & 0.55 & 0.48 & 0.45 & 0.75 & 17.14 & 1.11 & 0.39 & 0.19 & 0.81 & 0.49 & 0.77 & 22.23\\
				& ( 0.72,  0.20) & ( 0.15,  0.21) & ( 0.43,  0.18) & ( 4.52,  0.53) & ( 6.17,  0.75) & ( 0.76,  0.34) & ( 0.65,  0.17) & ( 0.28,  0.20) & ( 0.24,  0.31) & ( 0.16,  0.33) & ( 0.22,  0.23) & ( 0.50,  0.25) & ( 0.00, 17.14) & ( 0.89,  0.22) & ( 0.17,  0.22) & ( 0.09,  0.10) & ( 0.64,  0.17) & ( 0.25,  0.24) & ( 0.33,  0.44) & (17.16,  5.07)\\
				Soyo & 4.48 & 1.03 & 1.10 & 16.88 & 3.37 & 2.62 & 1.62 & 1.47 & 1.03 & 0.47 & 1.26 & 4.66 & 1.13 & 0.34 & 1.66 & 0.94 & 0.63 & 1.13 & 0.89 & 46.38\\
				& ( 4.22,  0.26) & ( 0.78,  0.25) & ( 0.60,  0.50) & (16.58,  0.30) & ( 2.97,  0.40) & ( 2.09,  0.54) & ( 0.83,  0.79) & ( 1.03,  0.44) & ( 0.70,  0.33) & ( 0.28,  0.19) & ( 0.81,  0.45) & ( 4.03,  0.63) & ( 0.68,  0.45) & ( 0.00,  0.34) & ( 1.10,  0.56) & ( 0.20,  0.74) & ( 0.36,  0.27) & ( 0.49,  0.64) & ( 0.46,  0.43) & (38.20,  8.18)\\
				Cattl & 0.70 & 1.23 & 1.99 & 0.99 & 0.88 & 1.74 & 0.38 & 1.54 & 1.10 & 0.89 & 0.82 & 0.67 & 0.87 & 1.85 & 1.16 & 1.25 & 0.87 & 0.99 & 0.64 & 19.39\\
				& ( 0.43,  0.27) & ( 0.66,  0.57) & ( 0.91,  1.08) & ( 0.52,  0.47) & ( 0.31,  0.56) & ( 0.57,  1.17) & ( 0.16,  0.22) & ( 1.22,  0.32) & ( 0.37,  0.73) & ( 0.58,  0.30) & ( 0.41,  0.42) & ( 0.38,  0.29) & ( 0.21,  0.66) & ( 1.22,  0.64) & ( 0.00,  1.16) & ( 0.85,  0.40) & ( 0.64,  0.23) & ( 0.32,  0.67) & ( 0.31,  0.33) & (10.07,  9.32)\\
				Milk & 1.76 & 0.84 & 2.16 & 0.40 & 1.00 & 0.76 & 0.63 & 0.64 & 1.44 & 1.06 & 0.56 & 0.84 & 0.36 & 0.99 & 2.01 & 0.20 & 0.26 & 1.44 & 0.30 & 17.46\\
				& ( 0.95,  0.81) & ( 0.27,  0.57) & ( 0.62,  1.55) & ( 0.17,  0.23) & ( 0.36,  0.64) & ( 0.35,  0.41) & ( 0.31,  0.32) & ( 0.36,  0.28) & ( 1.17,  0.27) & ( 0.52,  0.53) & ( 0.12,  0.45) & ( 0.42,  0.42) & ( 0.12,  0.25) & ( 0.26,  0.74) & ( 0.84,  1.17) & ( 0.00,  0.20) & ( 0.17,  0.09) & ( 0.49,  0.95) & ( 0.18,  0.13) & ( 7.65,  9.81)\\
				Lhog & 0.78 & 0.66 & 0.83 & 4.14 & 1.58 & 1.11 & 0.66 & 0.47 & 1.03 & 1.20 & 1.04 & 1.51 & 2.29 & 0.88 & 0.93 & 0.27 & 0.27 & 0.76 & 1.81 & 21.96\\
				& ( 0.33,  0.45) & ( 0.43,  0.24) & ( 0.44,  0.39) & ( 3.42,  0.73) & ( 0.56,  1.02) & ( 0.85,  0.26) & ( 0.22,  0.44) & ( 0.21,  0.27) & ( 0.62,  0.40) & ( 0.58,  0.62) & ( 0.49,  0.55) & ( 0.32,  1.19) & ( 0.66,  1.63) & ( 0.48,  0.40) & ( 0.63,  0.29) & ( 0.16,  0.10) & ( 0.00,  0.27) & ( 0.46,  0.31) & ( 1.01,  0.80) & (11.85, 10.11)\\
				Juice & 0.80 & 0.95 & 0.89 & 1.85 & 1.72 & 1.32 & 1.32 & 0.84 & 1.32 & 1.18 & 1.21 & 0.88 & 0.65 & 1.23 & 1.29 & 1.34 & 1.75 & 1.15 & 1.28 & 21.82\\
				& ( 0.20,  0.61) & ( 0.73,  0.22) & ( 0.53,  0.36) & ( 1.56,  0.29) & ( 1.04,  0.68) & ( 0.36,  0.96) & ( 0.24,  1.08) & ( 0.55,  0.29) & ( 0.80,  0.51) & ( 0.71,  0.46) & ( 0.69,  0.53) & ( 0.62,  0.26) & ( 0.29,  0.36) & ( 0.68,  0.55) & ( 0.32,  0.97) & ( 0.47,  0.88) & ( 0.46,  1.29) & ( 0.00,  1.15) & ( 1.02,  0.26) & (11.26, 10.57)\\
				Barl & 1.22 & 1.61 & 1.21 & 1.23 & 2.65 & 3.95 & 0.70 & 1.00 & 1.10 & 0.78 & 0.94 & 1.70 & 1.47 & 1.00 & 0.50 & 0.31 & 1.29 & 1.50 & 1.13 & 24.17\\
				& ( 0.68,  0.54) & ( 0.87,  0.74) & ( 0.43,  0.78) & ( 0.59,  0.64) & ( 1.52,  1.14) & ( 0.84,  3.11) & ( 0.21,  0.48) & ( 0.47,  0.53) & ( 0.29,  0.81) & ( 0.41,  0.37) & ( 0.33,  0.62) & ( 0.84,  0.86) & ( 0.35,  1.12) & ( 0.53,  0.47) & ( 0.30,  0.20) & ( 0.17,  0.15) & ( 1.00,  0.29) & ( 1.02,  0.49) & ( 0.00,  1.13) & (10.84, 13.34)\\\\
				\hline
				TO & 36.54 & 19.85 & 23.64 & 59.57 & 53.15 & 45.63 & 25.17 & 31.46 & 25.17 & 22.72 & 17.63 & 34.41 & 27.82 & 48.15 & 19.59 & 14.18 & 18.37 & 18.81 & 16.44 & 558.29\\
				& ( 27.33,   9.20) & ( 12.91,   6.94) & ( 13.15,  10.49) & ( 51.35,   8.21) & ( 41.20,  11.95) & ( 33.78,  11.85) & ( 16.68,   8.49) & ( 23.91,   7.55) & ( 17.78,   7.38) & ( 15.21,   7.51) & (  9.56,   8.06) & ( 24.03,  10.38) & ( 15.75,  12.07) & ( 38.16,   9.99) & (  9.45,  10.14) & (  6.89,   7.29) & ( 10.58,   7.79) & ( 10.17,   8.63) & ( 10.24,   6.20) & (388.16, 170.14)\\
				Inc.Own & 36.74 & 21.07 & 24.11 & 59.95 & 55.13 & 46.27 & 27.78 & 31.81 & 25.83 & 23.05 & 18.71 & 35.08 & 44.96 & 48.49 & 20.74 & 14.39 & 18.64 & 19.96 & 17.57 & TCI\\
				& (27.33,  9.41) & (12.91,  8.17) & (13.15, 10.96) & (51.35,  8.59) & (41.20, 13.93) & (33.78, 12.48) & (16.68, 11.10) & (23.91,  7.90) & (17.78,  8.05) & (15.21,  7.84) & ( 9.56,  9.15) & (24.03, 11.05) & (15.75, 29.21) & (38.16, 10.33) & ( 9.45, 11.29) & ( 6.89,  7.49) & (10.58,  8.06) & (10.17,  9.79) & (10.24,  7.34) & ($\rm TCI^C$, $\rm TCI^L$)\\
				NET & -0.22 & -1.30 & 1.57 & 2.85 & 7.01 & 3.51 & -1.93 & 0.12 & 0.67 & -0.20 & -3.68 & 1.66 & 5.59 & 1.77 & 0.20 & -3.27 & -3.59 & -3.01 & -7.73 & 29.38\\
				& ( 0.63, -0.86) & (-0.89, -0.41) & (-0.60,  2.18) & ( 5.96, -3.11) & ( 1.82,  5.18) & ( 1.18,  2.33) & (-0.74, -1.19) & ( 0.38, -0.25) & (-0.33,  1.00) & (-0.47,  0.27) & (-0.71, -2.96) & (-0.43,  2.10) & (-1.41,  7.00) & (-0.04,  1.81) & (-0.62,  0.82) & (-0.76, -2.52) & (-1.27, -2.32) & (-1.08, -1.93) & (-0.60, -7.13) & (20.43, 8.95)\\
				\bottomrule
		\end{tabular}}
		\begin{tablenotes}[para,flushleft]
			\tiny Notes: $R^{2}$ decomposed measures are based on a 200-day rolling-window VAR model with a lag length of order one (BIC). Values in parentheses represent contemporaneous and lagged effects, respectively.
		\end{tablenotes}
	\end{table}

	Additionally, corn is the primary net transmitter of shocks (7.01\%), followed by soybean meal  (5.59\%), and wheat (3.51\%). Several factors might make corn the main net contributor of shocks. First, the close relationship between corn and commodity markets has been widely reported \citep{kang2017dynamic}. Corn is one of the primary raw materials for biofuels, whose demand is closely linked to oil prices. Secondly, corn, requiring a significant amount of water, is particularly sensitive to climate change. 
	Third, corn plays a crucial role in the global food supply chain. Moreover, agricultural activities can also lead to significant carbon emission allowance. Similarly, both soybean meal and wheat are important agricultural products for many countries, which might make them risk contributors.
	
	Conversely, barley is the chief net receiver of shocks (-7.73\%), with cocoa  (-3.68\%) and leah hog (-3.59\%) following. Besides that, climate change is a net transmitter (1.57\%), while crude oil (-0.22\%) and carbon emission 
	allowance (-1.30\%) are net risk receivers. 
	In detail, carbon emission (2.12\%), cattle (1.99\%), and milk (2.16\%) receive the largest risk from climate change, while the largest spillovers to climate change are caused by crude oil (3.1\%). Barley receives the largest spillovers from wheat (3.95\%), with 0.84\% being caused by contemporaneous effects while 3.11\% originate from lagged spillovers. More specifically, corn and wheat receive major spillovers from each other. The spillover between crude oil and agricultural markets aligns with \citet{du2011speculation}.

	\subsection{Dynamic total connectedness}

\begin{figure}[H]
		\centering
		\includegraphics[width=1\linewidth]{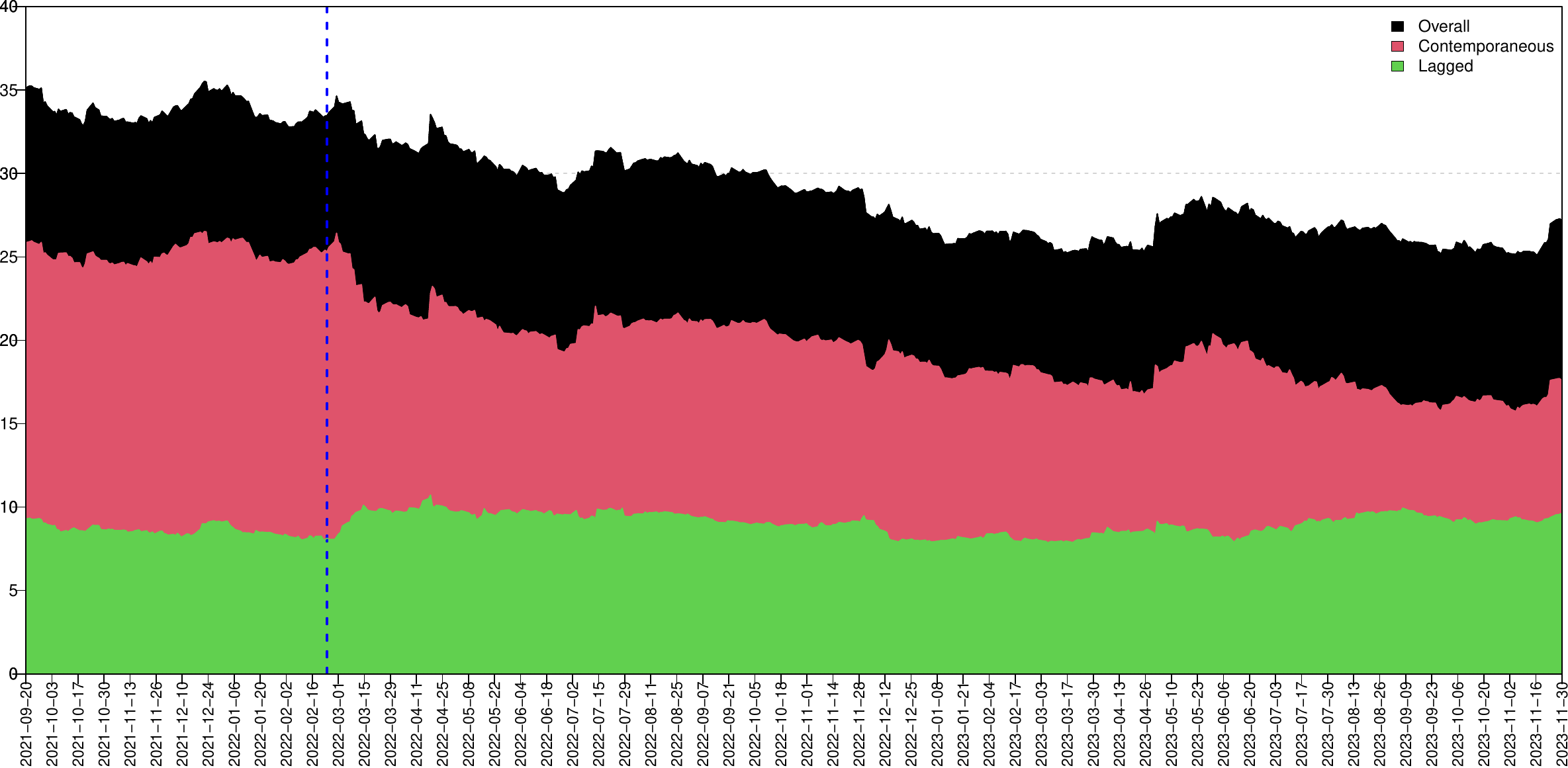}
		\caption{Dynamic total connectedness. The black margin visualizes the overall dynamic total connectedness while the dynamic contemporaneous and lagged connectedness are illustrated in red and green, respectively}
		\label{Fig:TCI:R2}
	\end{figure}

	Fig.~\ref{Fig:TCI:R2} presents the dynamic total connectedness. We note high market spillovers since 2021. The dynamic total connectedness decreased, indicating a diminishing COVID-19 impact, but rose sharply early in the Russia-Ukraine war. A decline followed, peaking at 28\% in June 2023. Then the TCI has plateaued at around 25\% throughout the investigation period. The lagged TCI stayed around 10\%, with an increase during the Russia-Ukraine war.
	
	Moreover, the overall and contemporaneous dynamic total connectedness measures behaved extremely similarly. Contemporaneous interdependencies dominate lagged ones, driving most total risk dynamics. The movement in the overall TCI is mainly driven by the contemporaneous dynamics and not by the lagged dynamics, which is in line with the  work of \citet{balli2023contemporaneous}. 
	
	\subsection{Dynamic FROM, TO and NET directional connectedness}
	
\begin{figure}[H]
		\centering
		\includegraphics[width=1.01\linewidth]{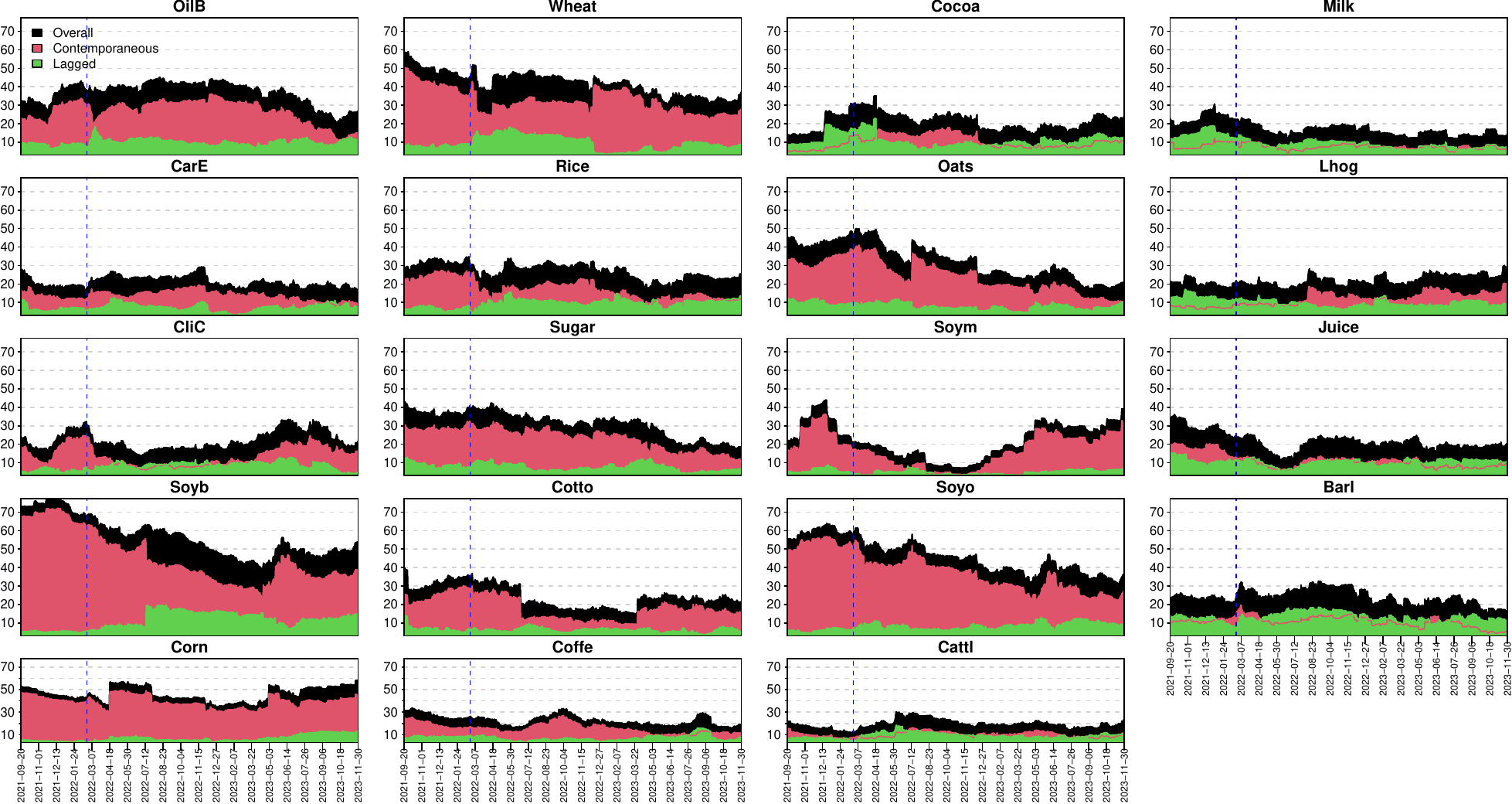}
		\caption{FROM connectedness.}
		\label{Fig:FROM:R2}
	\end{figure}

\begin{figure}[H]
		\centering
		\includegraphics[width=1.01\linewidth]{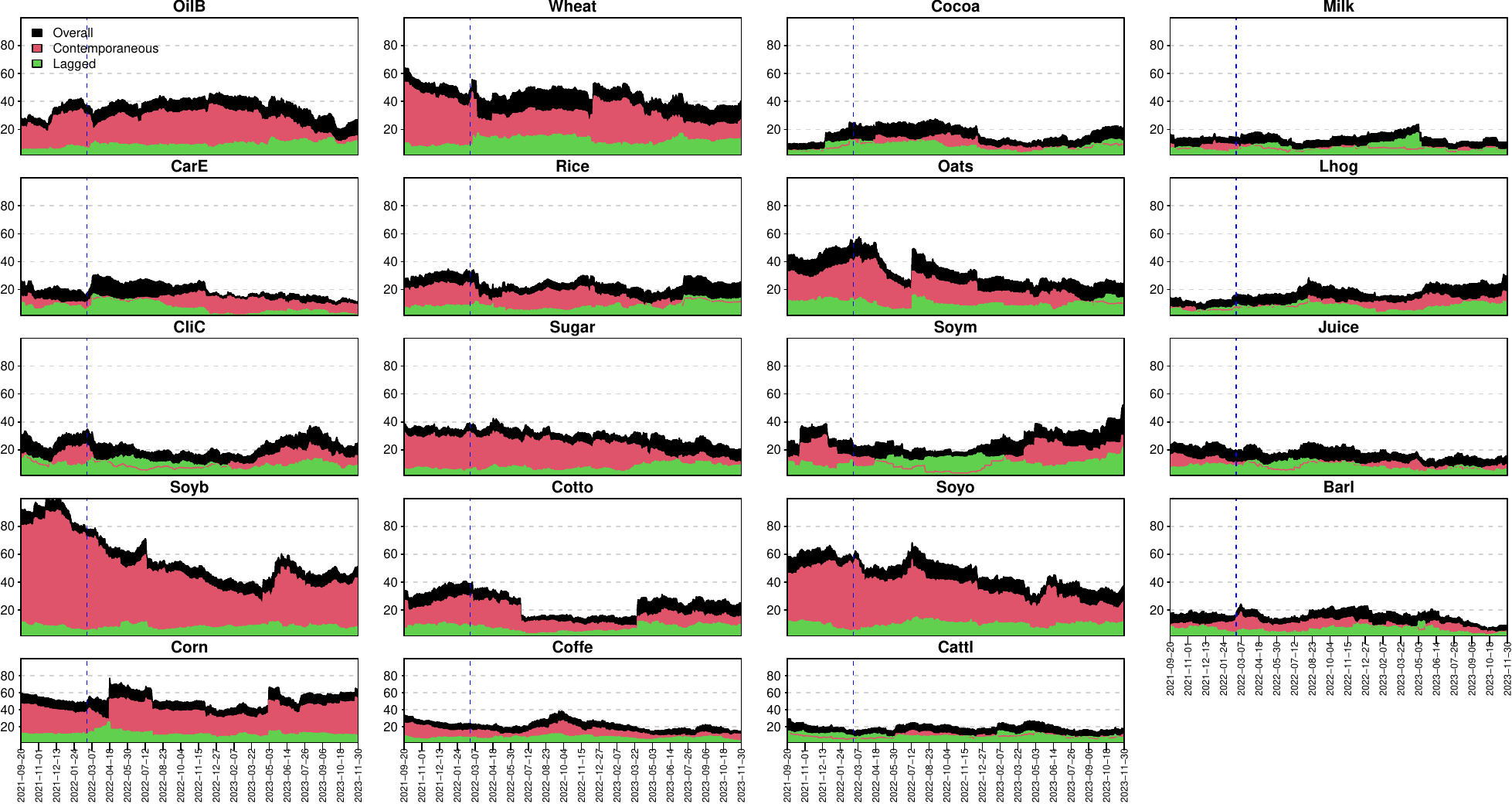}
		\caption{TO connectedness.}
		\label{Fig:TO:R2}
	\end{figure}

		Fig.~\ref{Fig:FROM:R2} and Fig.~\ref{Fig:TO:R2} illustrate the FROM and TO directional connectedness, respectively. The contemporaneous spillovers are more pronounced than the lagged spillovers. In Fig.~\ref{Fig:FROM:R2}, most variables' FROM connectedness fluctuated significantly during the Russia-Ukraine war, especially the crude oil, climate change, wheat and barley. Additionally, soybean has the highest degree of FROM connectedness, followed by soybean oil and wheat.

	Similarly, Fig.~\ref{Fig:TO:R2} shows that the TO directional connectedness for crude oil, carbon emission allowance, climate change, wheat, and barley shocked during the Russia-Ukraine war.  Besides that, the movement of the TO directional connectedness for carbon emission allowance, coffee, cocoa, cattle, milk, lean hog, juice, and barley remain below 30\%. Among them, the TO connectedness for cattle is relatively stable.

	\begin{figure}[H]
		\centering
		\includegraphics[width=1.01\linewidth]{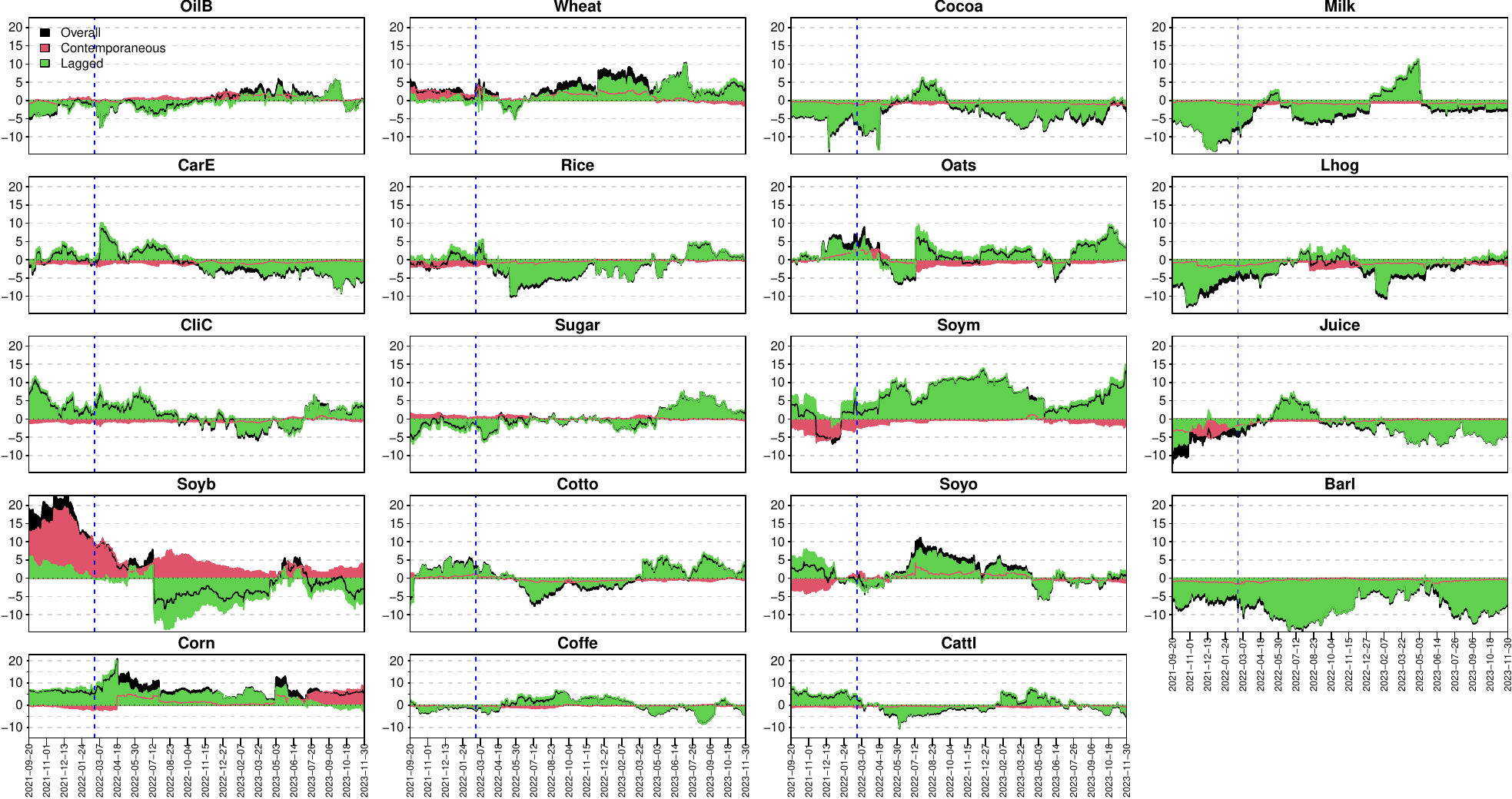}
		\caption{Net total directional connectedness.}
		\label{Fig:NET:R2}
	\end{figure}

	Fig.~\ref{Fig:NET:R2} presents that the net contributors of shocks mainly include climate change, corn, wheat, oats, soybean meat, and soybean oil, while cocoa, lean hog, juice, and barley mainly act as risk receivers. Among them, barley is the biggest receiver of net spillovers. The remaining markets alternate between being transmitters and receivers.
	
	A critical look suggests that crude oil shifts from a receiver to a transmitter since November 2022. However, carbon emission allowance shows an almost opposite pattern. Moreover, lagged net spillover of most markets exhibits higher similarity to overall net spillover effects, in line with \citet{liu2024decomposing}. Unlike other markets, soybean shows higher contemporaneous spillovers, followed by stronger lagged correlations.

	\subsection{Net pairwise directional connectedness measures}
	
\begin{landscape}
		\begin{figure}[htp]
			
			\centering
			\includegraphics[width=0.32\linewidth]{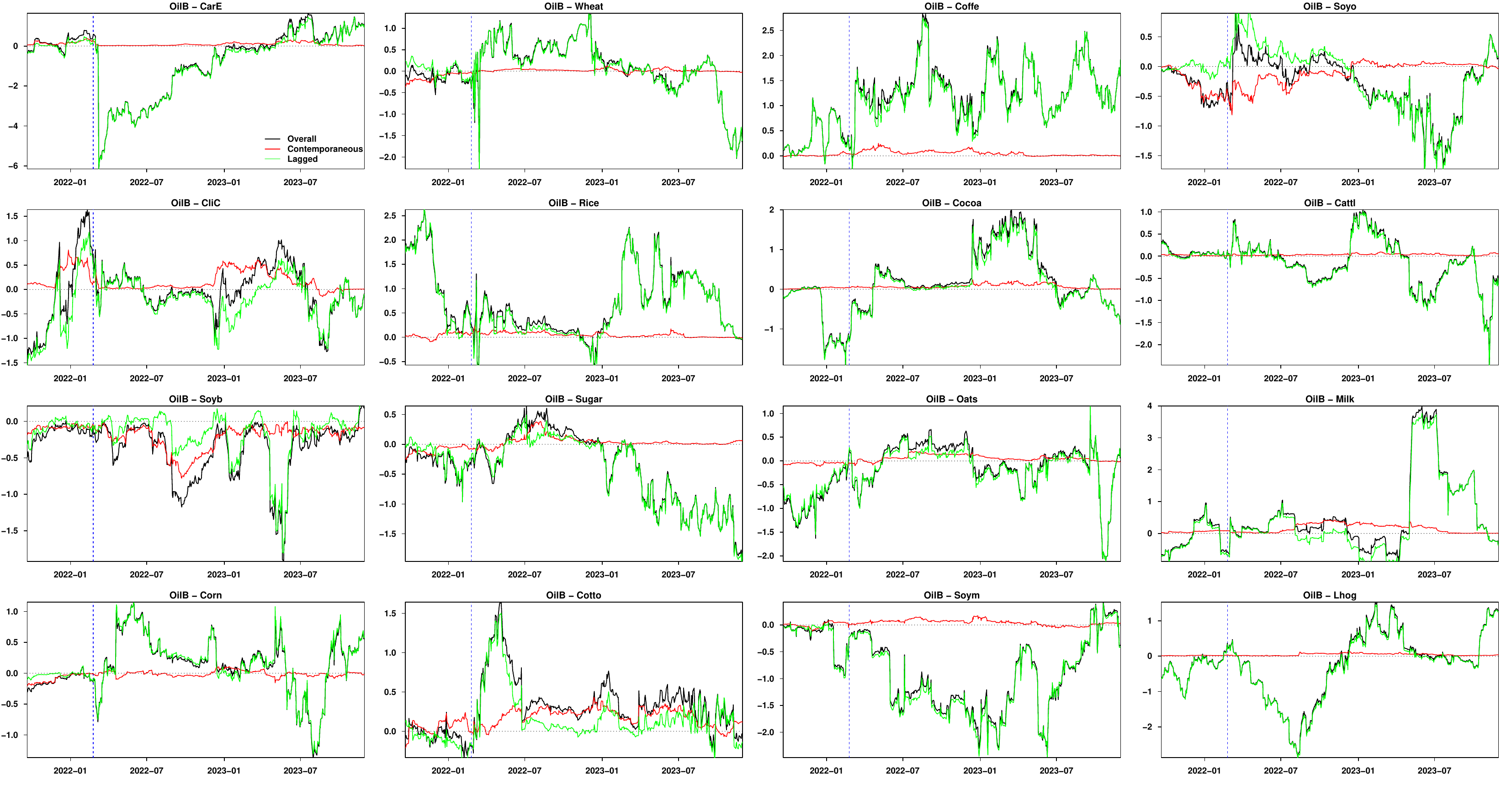}
			\includegraphics[width=0.32\linewidth]{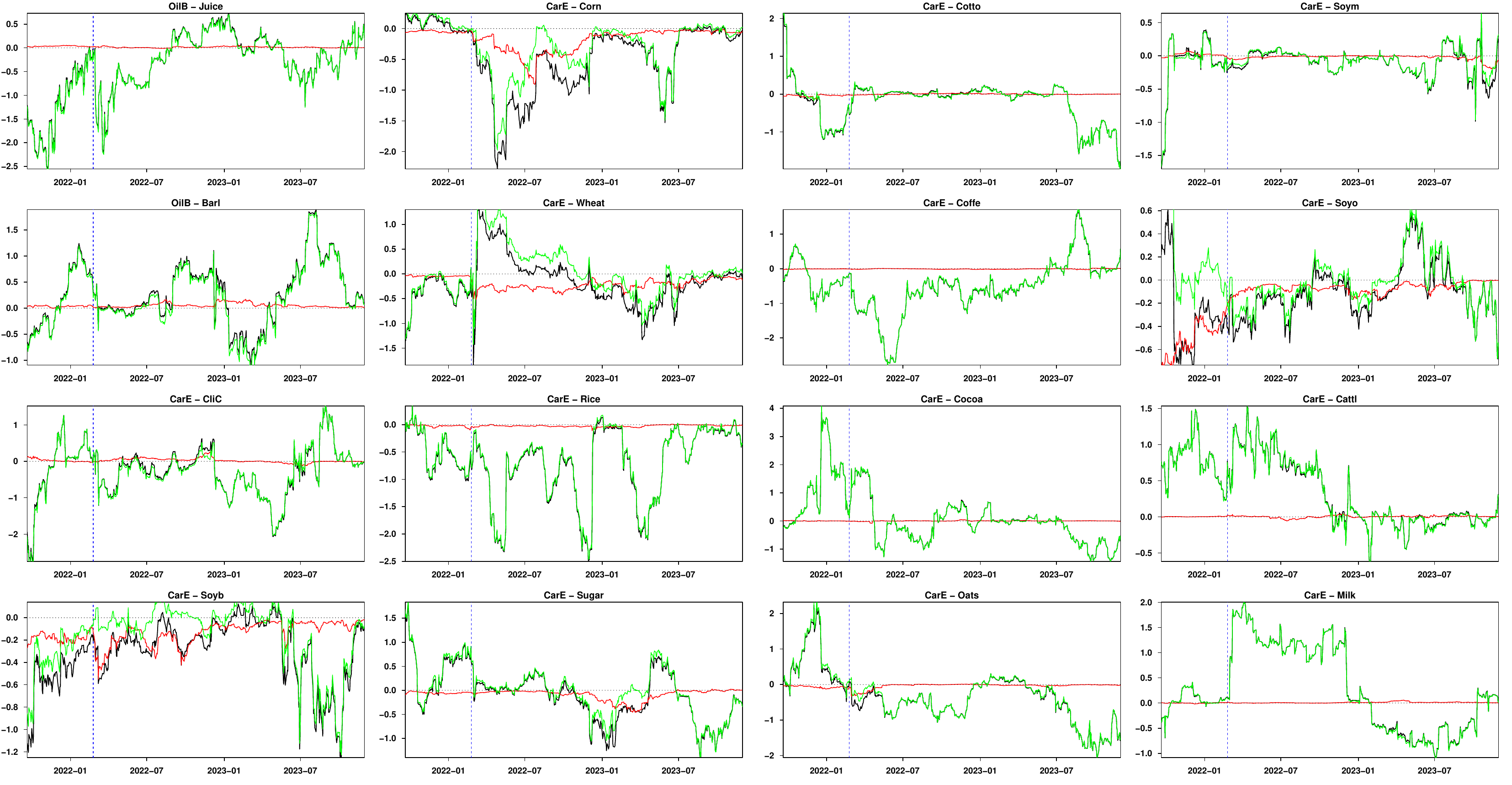}	 
			\includegraphics[width=0.32\linewidth]{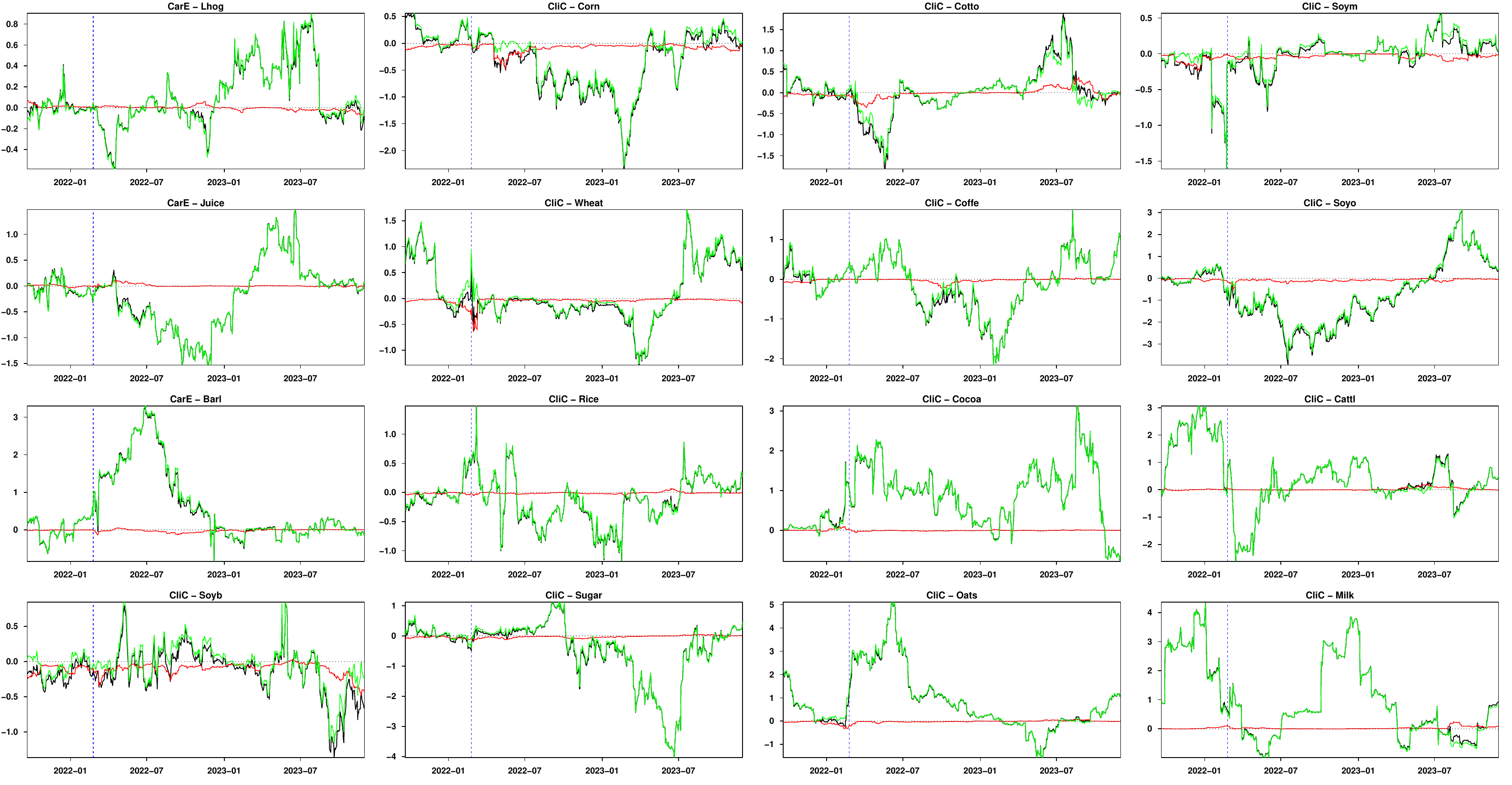}	\includegraphics[width=0.32\linewidth]{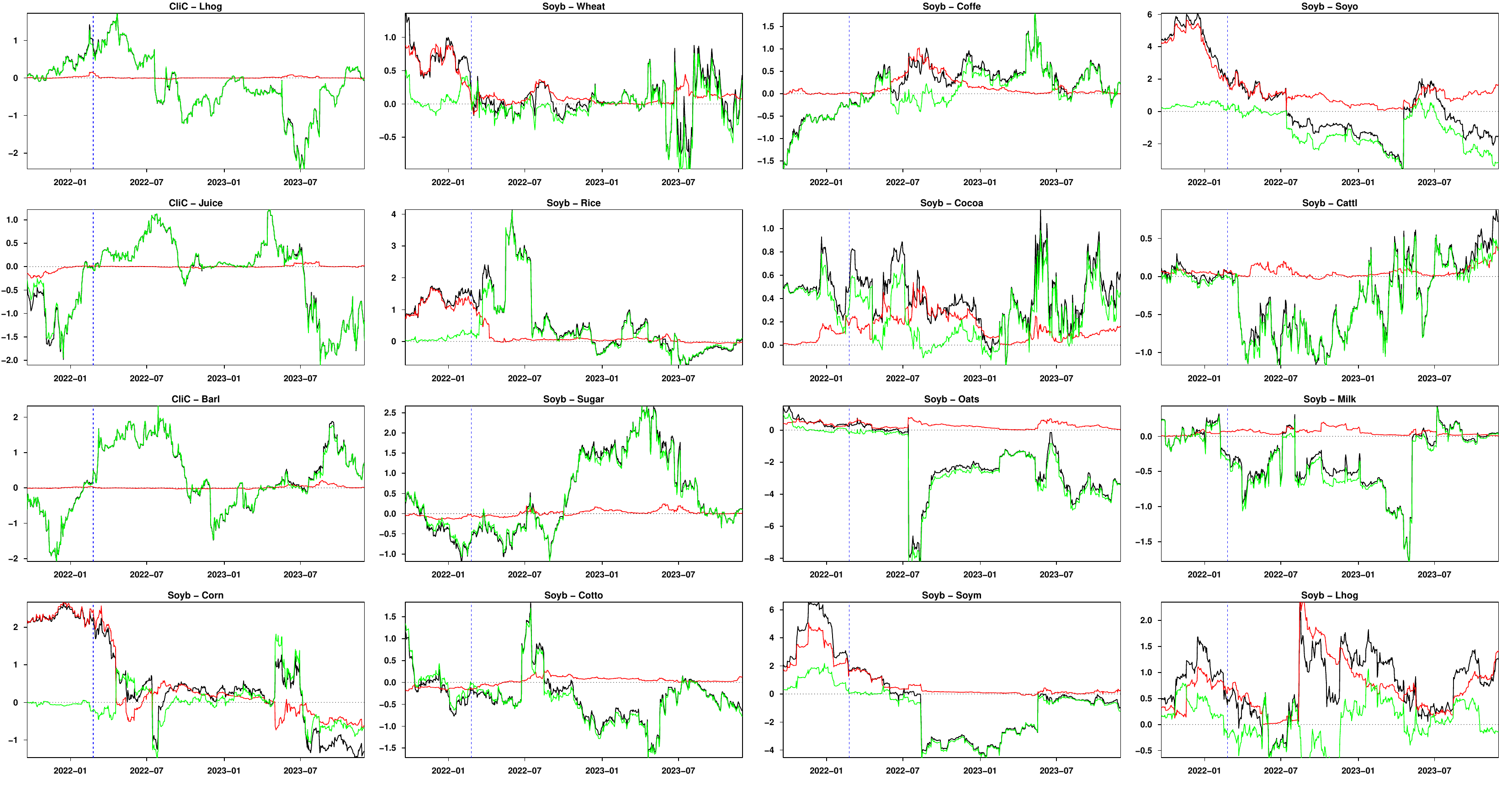}	\includegraphics[width=0.32\linewidth]{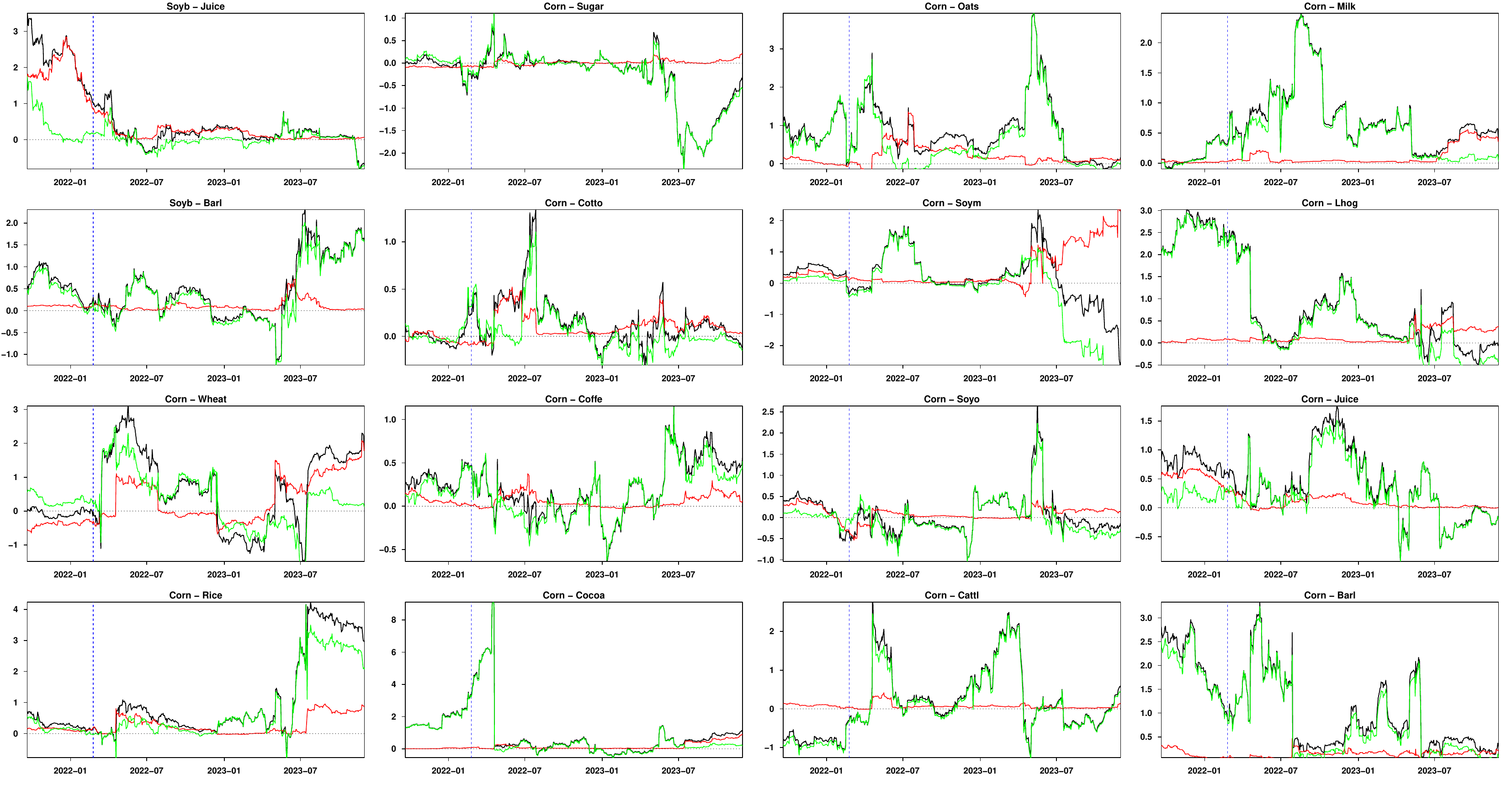}
			\includegraphics[width=0.32\linewidth]{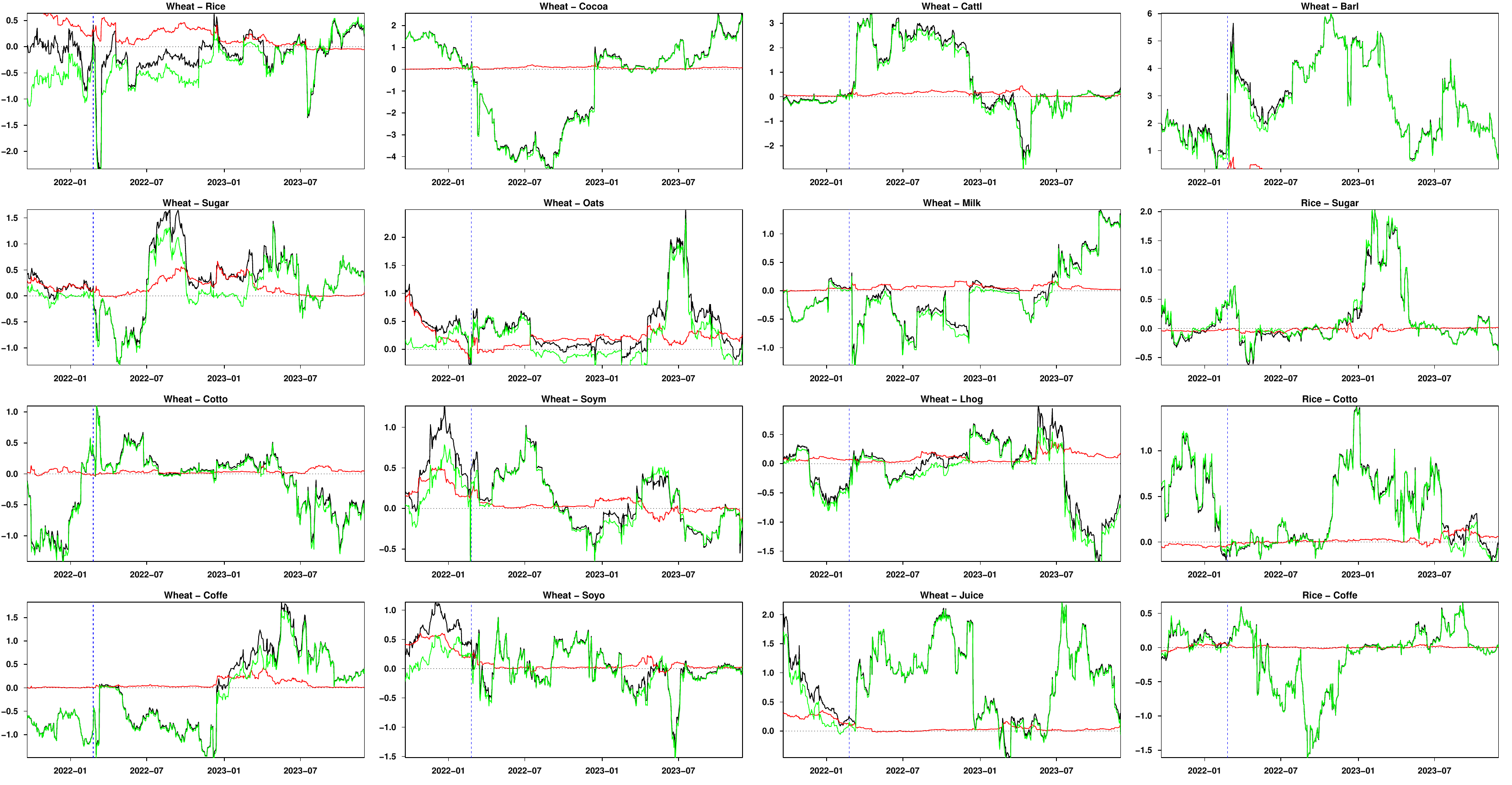}
			\includegraphics[width=0.32\linewidth]{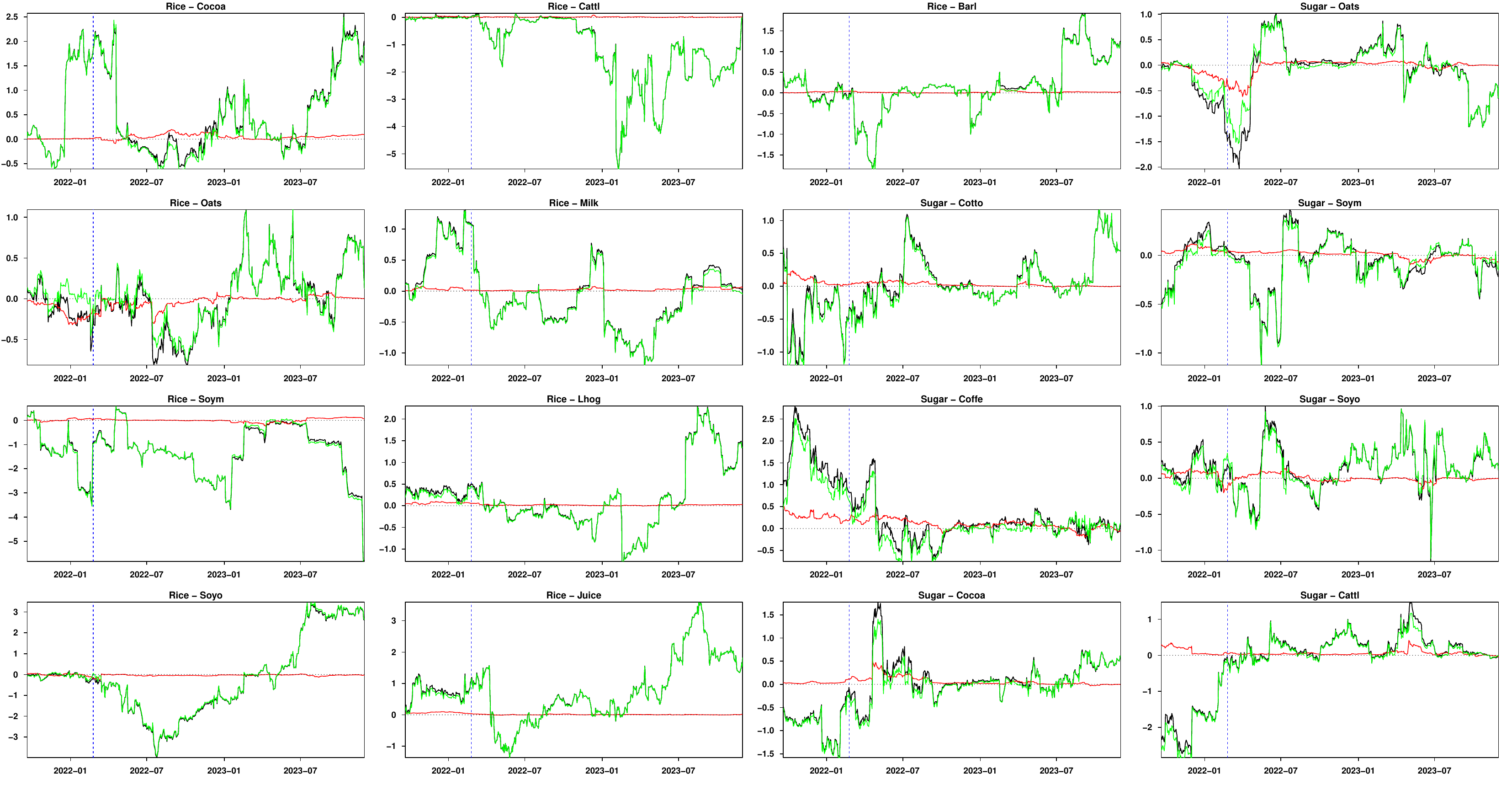}
			\includegraphics[width=0.32\linewidth]{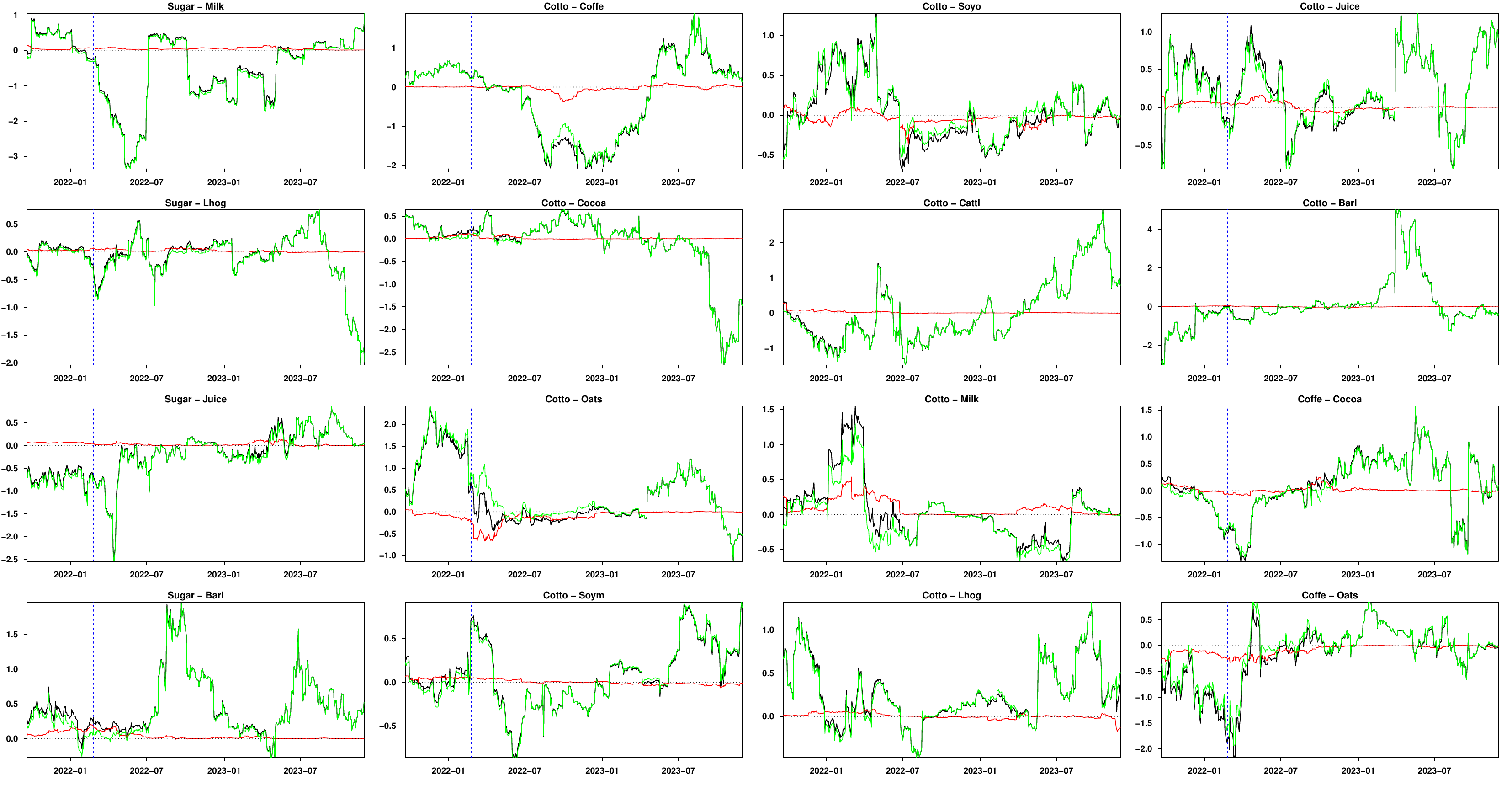}
			\includegraphics[width=0.32\linewidth]{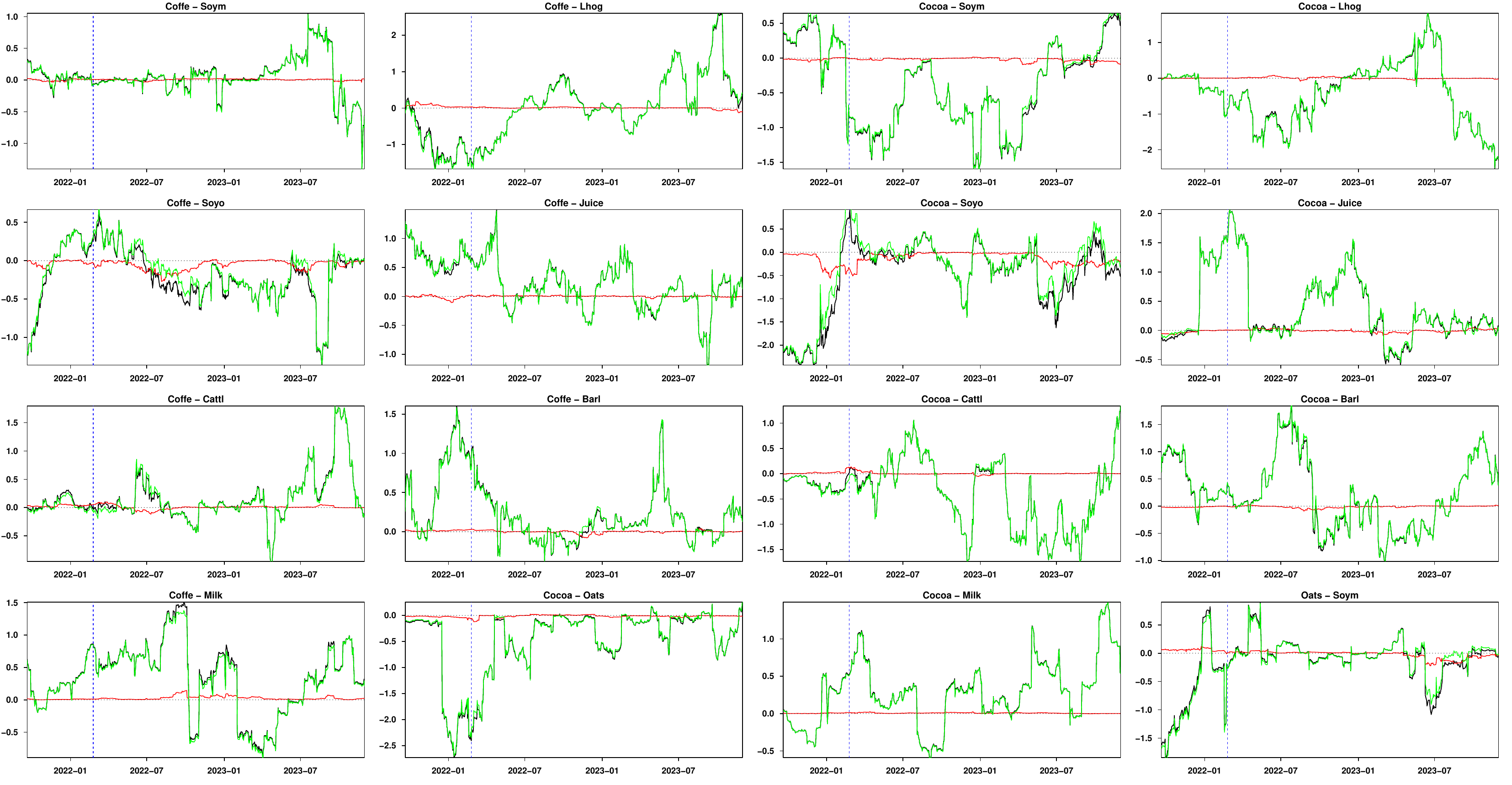}
			\includegraphics[width=0.32\linewidth]{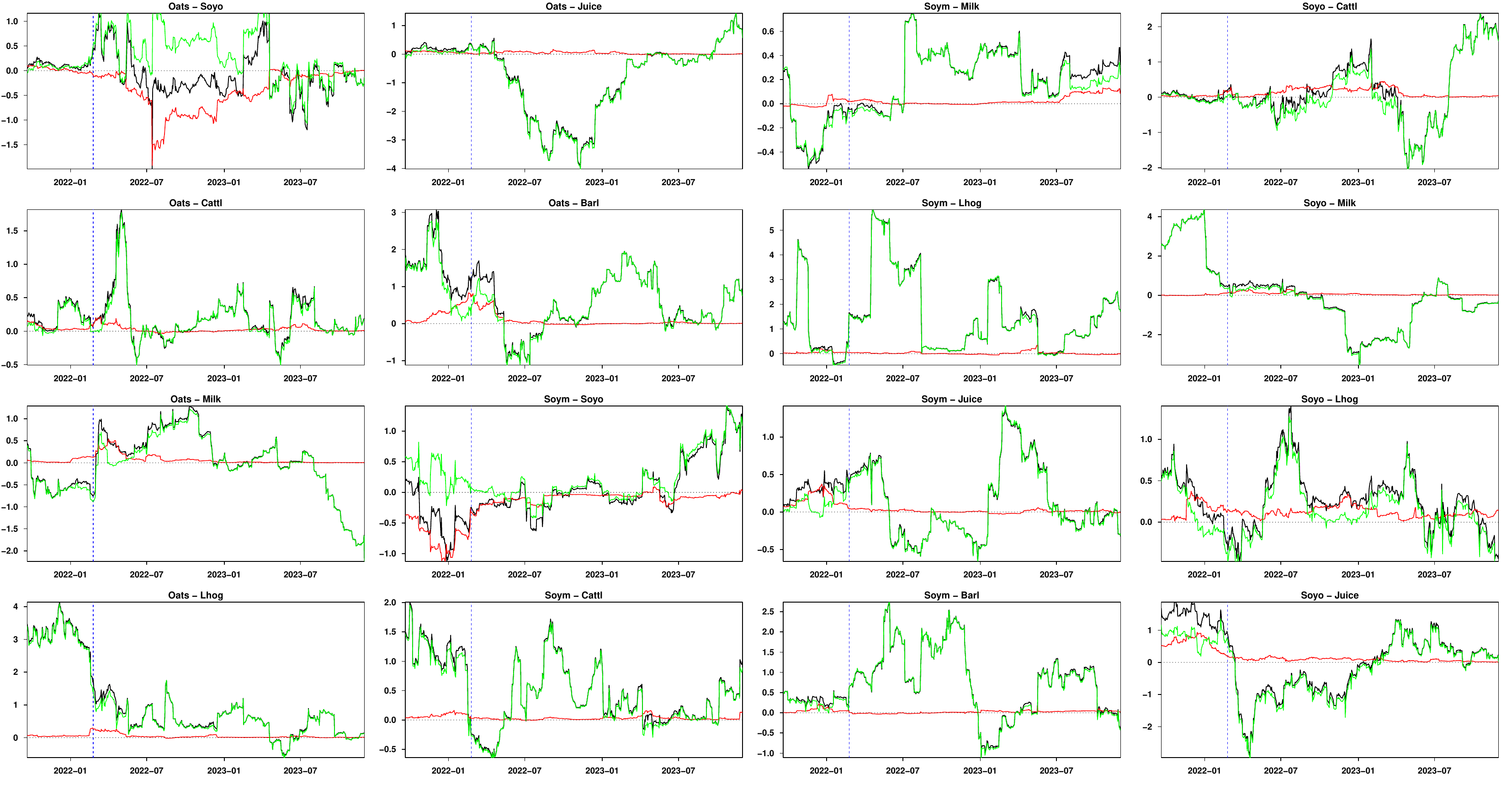}
			\includegraphics[width=0.32\linewidth]{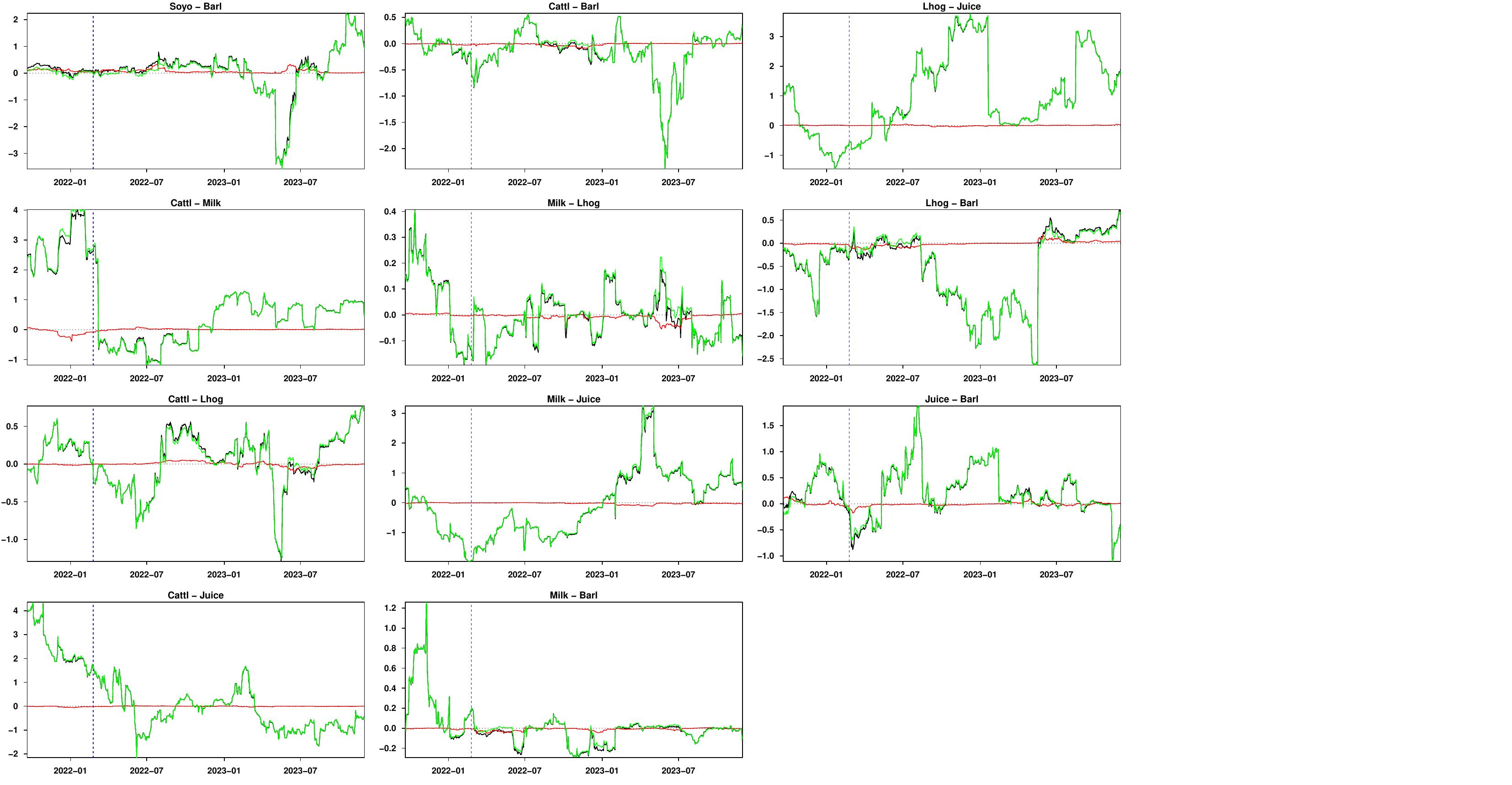}
			\caption{Net pairwise directional connectedness.}
			\label{Fig:NPDC:R2}
		\end{figure}
	\end{landscape}

	Fig.~\ref{Fig:NPDC:R2} shows the net pairwise risk spillovers of the system. A closer look reveals clear spillovers between climate change and other variables. Early in the sample period, climate risk is a net transmitter to the crude oil market,  consistent with \citet{jin2023geopolitical}. Cocoa, oats, and milk receive spillovers from climate change. Meanwhile, climate change receives spillovers from corn and soybean oil.
	
	Additionally, crude oil receives shocks from soybean and soybean meal, while contributing to rice, cotton, and coffee. However, this result differs from \citet{tiwari2022quantile}. And the net pairwise directional connectedness between crude oil and sugar switches its sign during the sample period, which corroborates the work of \citet{tiwari2022quantile}.

Moreover, lean hog is constantly dominated by soybean, corn, oats, soybean meal, and barley, which accords with Table~\ref{Tb:Average:Dynamic:Connectedness}. And barley receives high levels of spillovers from carbon emission allowance, wheat, and corn. Meanwhile, corn and rice transfer risks to the carbon market. We also observe spillovers from cattle, soybean, and soybean meal to rice.

	\subsection{Net pairwise network analysis}
	
\begin{figure}[H]
		\centering
		\includegraphics[width=0.7\linewidth]{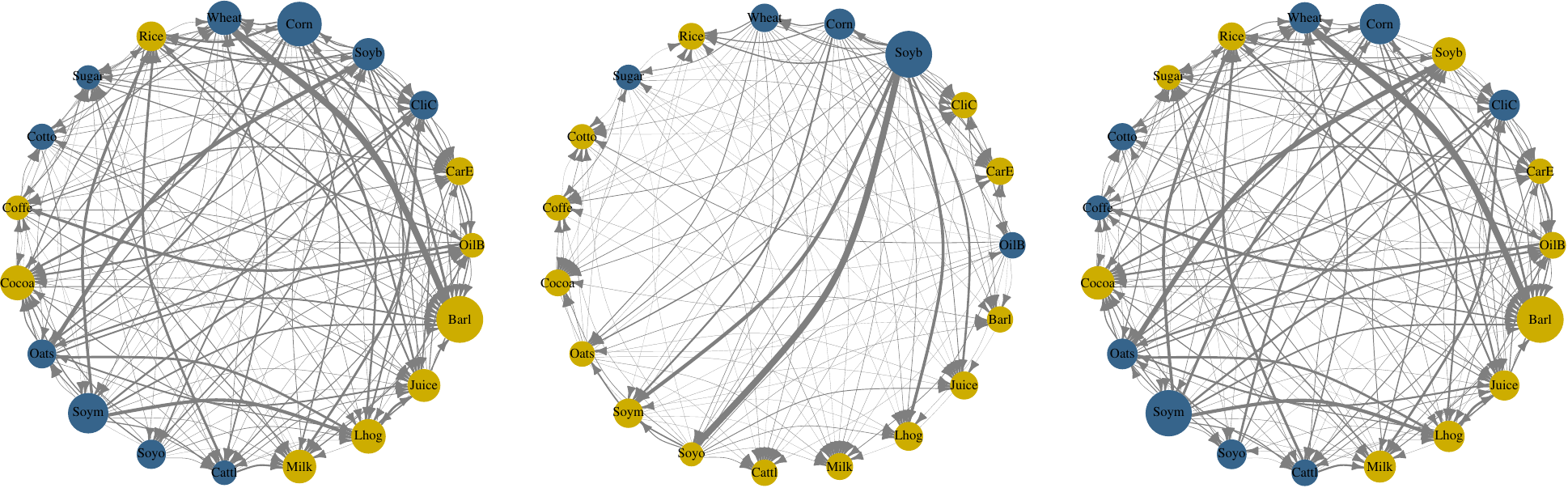}
		\caption{Network connectedness. From left to right, they correspond respectively to overall, contemporaneous, and lagged average network dependency.}
		\label{Fig:Network:R2}
	\end{figure}

	Fig.~\ref{Fig:Network:R2} illustrates the net pairwise directional connectedness measures, with blue dots for transmitters and yellow dots for receivers. Carbon emission allowance, barley, and rice exhibit negative net spillover effects. Corn and wheat have notable positive risk spillover effects, which is in line with \citet{guo2023climate}. Moreover, there is a high level of overall and lagged connectedness in wheat and barley, aligning with \citet{banerjee2024volatility}. 
	
	\subsection{Robustness check}	
	
\begin{figure}[H]
		\centering
		\includegraphics[width=0.6\linewidth]{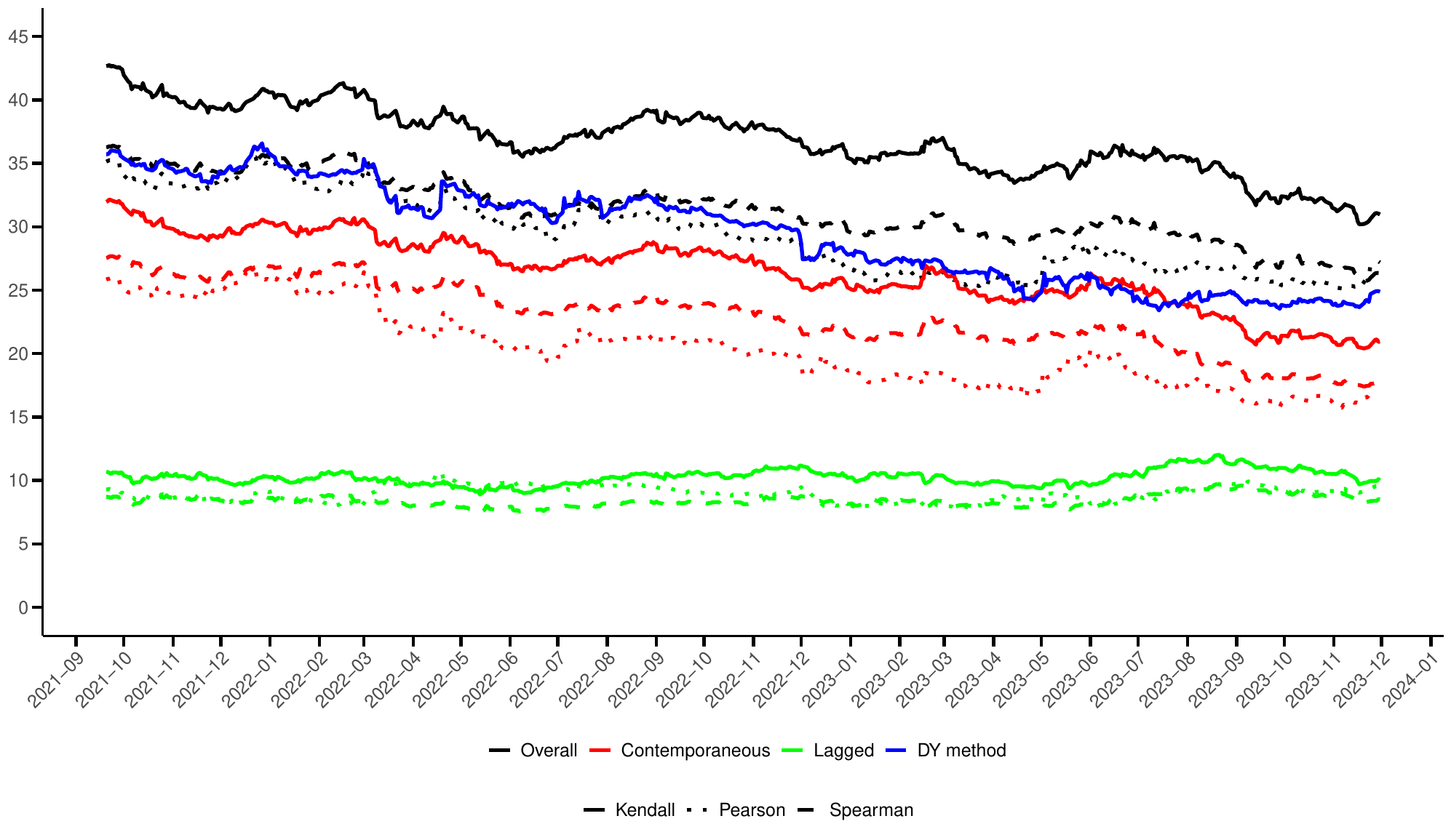}
		\caption{Robustness check: Dynamic total connectedness. The black line depicts the overall dynamic total connectedness while the dynamic contemporaneous and lagged connectedness are illustrated in red and green, respectively. The three distinct line types represent three different correlation coefficients: Kendall's $\tau$, Spearman's $\rho$, and Pearson's $r$. The blue line illustrates the DY method.}
		\label{Fig:Robustness:R2}
	\end{figure}

	Fig.~\ref{Fig:Robustness:R2} illustrates the robustness checks based on the dynamic total connectedness. We replace Pearson with Spearman and Kendall correlation coefficients and use the DY method as an alternative to verify TCI robustness. The dynamic TCIs of the three correlation types are similar, indicating robust results.
	
	\section{Conclusion}
	\label{S5:conclusion}
	
	This paper utilizes a novel $R^{2}$ decomposed connectedness approach to identify the contemporaneous and lagged effects across futures markets of agriculture, crude oil, carbon emission allowance, and climate change. The analysis shows the total TCI is driven by more volatile contemporaneous dynamics than lagged ones. Moreover, the total connectedness has increased due to the Russia-Ukrainian war. Additionally, there exist heterogenous spillover effects among agricultural markets. In detail, corn, soybean meal and wheat are the primary net transmitters, while barley, cocoa and lean hog act as main risk receivers. And we observe a degree of spillover from climate change to other markets. As our study focuses on the current factors influencing agricultural market spillovers, future research could explore the impact of fintech and other financial innovations on agricultural market risk transmission.
	
	Our results have the following implications: Given that contemporaneous spillover effects are more significant than lagged ones, policymakers and investors should focus on real-time monitoring and swift response mechanisms to manage risks as they occur. Moreover, they should implement differentiated risk management measures for various agricultural products, especially key commodities. In detail, hedging strategies could be more aggressive for contributors, while protective measures might be prioritized for receivers. Additionally, policymakers should invest in risk-resilient agricultural practices, strengthening responses to multiple risk sources.

	\section*{Acknowledgements}

	The authors would like to acknowledge the support of the Shanghai Planning Office of Philosophy and Social Science (2022EJB006).

\end{document}